\documentclass[11pt,titlepage,a4paper]{myarticle}

\usepackage[T1]{fontenc} 
\usepackage{comment} 
\RequirePackage[colorlinks=true
,urlcolor=blue
,anchorcolor=blue
,citecolor=blue
,filecolor=blue
,linkcolor=blue
,menucolor=blue
,linktocpage=true
,pdfproducer=medialab
,pdfa=true
]{hyperref}
\usepackage{amsmath, mathrsfs, amsfonts, amssymb, amsthm, mathtools, graphicx, color, ucs, xparse, tikz, lmodern, physics, varioref, tensor, cite}

\renewcommand{\dfrac}{\displaystyle\frac}

\newcommand{\Z}{\mathbb{Z}}

\newcommand{\8}{{\S^1\vee \S^1}}

\newcommand\mperiod[1][\rlap]{#1{\ .}}	
\newcommand\mcomma[1][\rlap]{#1{\ ,}}

\def\S{{\bf {S}}}

\def\Z{{\bf {Z}}}

\newenvironment{eq}
    {\begin{equation}
    \begin{aligned}
    }
    { 
    \end{aligned}
    \end{equation}
    \ignorespacesafterend
    }

\begin{document} 

\preprint{ {\tt IFT-UAM/CSIC-26-53}\\	}

\title{Heterotic Ouroboros}
\author{Chiara Altavista${}^1$, Salvatore Raucci${}^{1,2}$, Angel M.~Uranga${}^1$, Chuying Wang${}^1$
     \oneaddress{
     ${}^1$Instituto de F\'{i}sica Te\'{o}rica IFT-UAM/CSIC, C/ Nicol\'{a}s Cabrera 13--15, Campus de Cantoblanco, 28049 Madrid, Spain \\
     ${}^2$Departamento de F\'{i}sica Te\'{o}rica, Universidad Aut\'{o}noma de Madrid, Cantoblanco, 28049 Madrid, Spain\\ 
      {~}\\
      \email{chiara.altavista@estudiante.uam.es, salvatore.raucci@uam.es, angel.uranga@csic.es, chuying.wang@ift.csic.es}
}}

\Abstract{ \small
M-theory on ${\mathbf{S}}^1\vee{\mathbf{S}}^1$ has recently been proposed to yield, via quotients, ten-dimensional non-supersymmetric string theories. We revisit the construction that leads to the heterotic theories, finding a consistent set of rules that reproduces the light spectra and gauge groups, including indications of their global structure. Our approach uses the gauge enhancement mechanism of type I' string theory, applied to a setting in which the type I' interval is curled onto itself, with its two boundaries separated by a branch cut. Using these tools, we reproduce the ten-dimensional heterotic theories and provide some evidence for junctions among them.
}

\maketitle
\setcounter{page}{1}

\setcounter{tocdepth}{2}
\tableofcontents

\section{Introduction}

String dualities point to a picture in which string theories are local descriptions of an underlying quantum theory, M-theory. Perturbative strings then serve as local charts, and string dualities as transition maps. However, our knowledge of this paradigm is largely limited to supersymmetric constructions and confined to \emph{geometric} sectors of M-theory.

Recently, a new approach to non-supersymmetric type 0 theories has been proposed in \cite{Baykara:2026gem} in terms of M-theory on $\8$, two circles joining at a point, interpreted as a `quantum geometry'. M-theory on this space is defined by the boundary conditions assigned to all the 11d fields, shifting the focus from smooth compactifications to functions over (topological) spaces. A possible physical interpretation is that there are two separate components of the moduli space of M-theory. In the first, M-theory is formulated in geometric terms, thus giving rise to the well-known compactifications and dualities of string theories. The second component is not geometric and leads, in the proposal of \cite{Baykara:2026gem}, to non-supersymmetric string theories.

This approach has been further explored in \cite{Altavista:2026evd,Baykara:2026vdc}, with the inclusion of type 0 orientifolds from $\Z_2$ quotients of $\8$. In \cite{Baykara:2026vdc}, the authors also proposed a connection with the non-supersymmetric ten-dimensional heterotic theories by taking the $\Z_2$ quotient of $\8$ that exchanges the two $\S^1$'s. It was argued that this action could reproduce the E-type heterotic strings (together with the D-type heterotics after further operations), although no precise microscopic or `quantum-geometric' understanding was provided. The aim of this work is to build on this proposal and provide a string-inspired `geometric' interpretation for the different heterotic strings from M-theory on $\8$.

There are seven 10d non-supersymmetric heterotic string theories\footnote{See, e.g., \cite{Mourad:2017rrl,Basile:2021vxh,Angelantonj:2024tns,Raucci:2024fnp,Leone:2025mwo,Dudas:2025ubq} for recent reviews of non-supersymmetric strings.}. These were initially found by constructing all possible modular-invariant partition functions \cite{Alvarez-Gaume:1986ghj,Dixon:1986iz,Kawai:1986vd}, and the classification was recently shown to be complete in \cite{BoyleSmith:2023xkd} using modern CFT techniques. These heterotic theories can be separated into two groups, the E- and the D-type heterotics, according to whether orbifold operations connect them to the supersymmetric $E_8\times E_8$ or $Spin(32)/\Z_2$ theories. The non-supersymmetric and tachyon-free $SO(16)\times SO(16)$ theory~\cite{Alvarez-Gaume:1986ghj} is special in that it belongs to both---it has a D-type gauge group and matter fields in spinor representations. The various heterotic theories and their light (massless and tachyonic) spectra are shown in Table \ref{table:heterotics}. 

\begin{table}[h!]
    \centering
    \resizebox{\textwidth}{!}{\begin{tabular}{|c|c|c|c|c|c|}\hline
    \multicolumn{3}{|c|}{E-type} & \multicolumn{3}{|c|}{D-type} \\\hline
         Gauge group & Tachyons & Fermions & Gauge group & Tachyons & Fermions \\\hline  
         $E_8\times SO(16)$ & $(\mathbf{1,16})$ &
         $\begin{matrix}
             (\mathbf{1,128})_+\\
             (\mathbf{1,128'})_- 
         \end{matrix}$ & $SO(32)$ & $(\mathbf{32})$ & --- \\\hline
         $[E_7\times SU(2)]^2$ & $(\mathbf{1,2,1,2})$ &
         $\begin{matrix}
             (\mathbf{56,2,1,1})_+\\
             (\mathbf{1,1,56,2})_+\\
             (\mathbf{56,1,1,2})_-\\
             (\mathbf{1,2,56,1})_-
         \end{matrix}$ & 
         $SO(24)\times SO(8)$ & $(\mathbf{1,8})$ &
         $\begin{matrix}
             (\mathbf{24,8_v})_+\\
             (\mathbf{24,8_s})_-
         \end{matrix}$ \\\hline
         $(E_8)_2$ & $(\mathbf{1})$ &
         $\begin{matrix}
             (\mathbf{248})_+\\
             (\mathbf{248})_-
         \end{matrix}$ & $SU(16)\times U(1)$ & $(\mathbf{1},\pm 1)$ &
         $\begin{matrix}
             (\mathbf{120},+2)_+\\
             (\mathbf{\overline{120}},-2)_+\\
             (\mathbf{120},-2)_-\\
             (\mathbf{\overline{120}},+2)_-
         \end{matrix}$ \\\hline
         $SO(16)\times SO(16)$ & --- &
         $\begin{matrix}
             (\mathbf{128,1})_+\\
             (\mathbf{1,128})_+\\
             (\mathbf{16,16})_-
         \end{matrix}$ & $SO(16)\times SO(16)$ & --- &
         $\begin{matrix}
             (\mathbf{128,1})_+\\
             (\mathbf{1,128})_+\\
             (\mathbf{16,16})_-
         \end{matrix}$ \\\hline
    \end{tabular}}
    \caption{Massless and tachyonic field content of the seven 10d non-supersymmetric heterotic strings, organized into E- and D-type classes (adapted from \cite{Baykara:2026vdc}). Theories within each class are ordered by decreasing numbers of tachyons. The $SO(16)\times SO(16)$ is repeated because it can be regarded as belonging to both classes. The E- and D-type theories in the same row are related as explained in section \ref{sec:d-limit}. }
    \label{table:heterotics}
\end{table}

The global structures of the gauge groups of the seven non-supersymmetric heterotic theories were recently determined in \cite{Fraiman:2023cpa}.\footnote{See also \cite{McInnes:1999va,McInnes:1999pt,Basile:2023knk} for the $SO(16)\times SO(16)$ theory.} The groups are
\begin{eq}\label{global}
    & E_8\times Spin(16) \mcomma \qquad \dfrac{[E_7\times SU(2)]^2}{\Z_2}\rtimes \Z_2 \mcomma \qquad  (E_8)_2 \mcomma \qquad \frac{Spin(16)\times Spin(16)}{\Z_2}\rtimes \Z_2 \mcomma  \\
        &Spin(32) \mcomma \qquad \frac{Spin(24)\times Spin(8)}{\Z_2}  \mcomma \qquad  \frac{SU(16)}{\Z_2}\times U(1) \rtimes \Z_2 \mperiod
\end{eq}
In the following, we will continue using the familiar terminology, thus referring to them by their Lie algebras. The global structure of the groups will be important in later sections.

There have been several attempts to obtain string dualities for non-supersymmetric heterotic strings; see, e.g., \cite{Blum:1997cs,Blum:1997gw,Blumenhagen:1999ad,Faraggi:2007tj,Acharya:2022shu,Larotonda:2024thv,Fraiman:2025yrx}, but the proposal of~\cite{Baykara:2026vdc} has the virtue of comprising all 10d cases---tachyonic and not---in a single framework, generalizing the picture of string theories as local charts in the full M-theory at the cost of introducing non-geometric and singular configurations.

In this work, we provide rules that resolve the singular configuration of coincident boundaries and identification points from the proposal of~\cite{Baykara:2026vdc} for heterotic strings, gaining insight into the quantum geometry of quotients of $\8$. In particular, we explore the peculiar features that emerge in the presence of boundaries, finding that a surprising notion of locality emerges, related to the quantum identification point.

Our approach relies on compactifying on an additional $\S^1$ to reach a type IIA description. The heterotic construction maps to type IIA on a $\Z_2$ quotient of $\8$, namely a quantum version of type I', curled onto itself so that the two boundaries are `coincident'. We call this setup the \emph{ouroboros} configuration. This provides a more solid base on which we can formulate the rules governing the setup. In particular, our rules rely on the behavior of D8-branes in the neighborhood of O8-planes and their strong coupling limits, suitably modified to take into account the quantum nature of the configuration. In this way, it is possible to reproduce the gauge groups of the non-supersymmetric heterotic theories, with suggestive hints on their global structures, and the light spectra (tachyonic and massless) of both E-type and D-type heterotics. The latter are obtained by a particular limit in which the ouroboros shrinks, as argued in~\cite{Baykara:2026vdc}.

This realization of non-supersymmetric heterotic strings also provides a quantum-geometric realization of the string junctions of~\cite{Altavista:2026edv,Tachikawa:2026jaj}.
The picture that we develop through type I' string theory allows us to interpret some of the junctions in a quantum-geometric fashion and is a powerful tool to generalize the construction to less simple cases.

This work is organized as follows. In section~\ref{sec:review}, we briefly review the picture of~\cite{Baykara:2026gem} for M-theory on $\8$ and the approach to heterotic strings of~\cite{Baykara:2026vdc}. In section~\ref{sec:e-limit}, we analyze the quantum-geometric realization of the E-type heterotics. We start by presenting the general strategy of our approach---based on type I'-like configurations---in section~\ref{sec:e-strategy}, and then we address the microscopic details of the E-type limits in section~\ref{sec:e-microscopic}, presenting the rules that the non-geometric setup must satisfy. In section~\ref{sec:e-heterotics}, we apply the rules to obtain all the E-type heterotic theories from M-theory on the quantum geometry.
Section~\ref{sec:d-limit} deals with the D-type heterotic theories, following an analogous path: we outline the general strategy in section~\ref{sec:d-strategy}, present the rules of the relevant limit in section~\ref{sec:d-rules}, and apply them to obtain the D-type heterotics in section~\ref{sec:d-heterotics}.
In section~\ref{sec:junctions}, we provide some evidence that the tools we developed in previous sections can realize junctions of 10d heterotic string theories that were recently introduced in~\cite{Altavista:2026edv,Tachikawa:2026jaj}. We set up the formalism in section~\ref{sec:junctions-ide} and present some examples in sections~\ref{sec:junctions-e-examples} and~\ref{sec:junctions-d-examples}.
Appendix~\ref{app:enhancement} contains some details on the type I' enhancements that are crucial for the analysis in section~\ref{sec:e-microscopic}. 

\section{Non-supersymmetric heterotics and M-theory on \texorpdfstring{$\8$}{S1 V S1}}
\label{sec:review}

In this section, we review the relation proposed in \cite{Baykara:2026vdc} between M-theory on the quotient of $\8$ by the exchange of the two $\S^1$s and the E-type heterotic theories, and the subsequent map to the D-type theories.

M-theory on a quantum version of the wedge sum $\8$ (two $\S^1$'s joining at a point) corresponds to 10d type 0A string theory~\cite{Baykara:2026gem}. The presence of two circles explains the doubling of the RR fields with respect to the standard compactification of M-theory on $\S^1$ that gives rise to type IIA string theory. A `quantum' nature of $\8$ is required because the joining point must actually be a superposition of all the possible pairs of points in the two circles, so that the configuration is symmetric under translation in both $\S^1$'s; this corresponds to the gauge invariance of the two type 0A RR 1-forms. 

\begin{figure}[htb]
\begin{center}
\includegraphics[scale=.35]{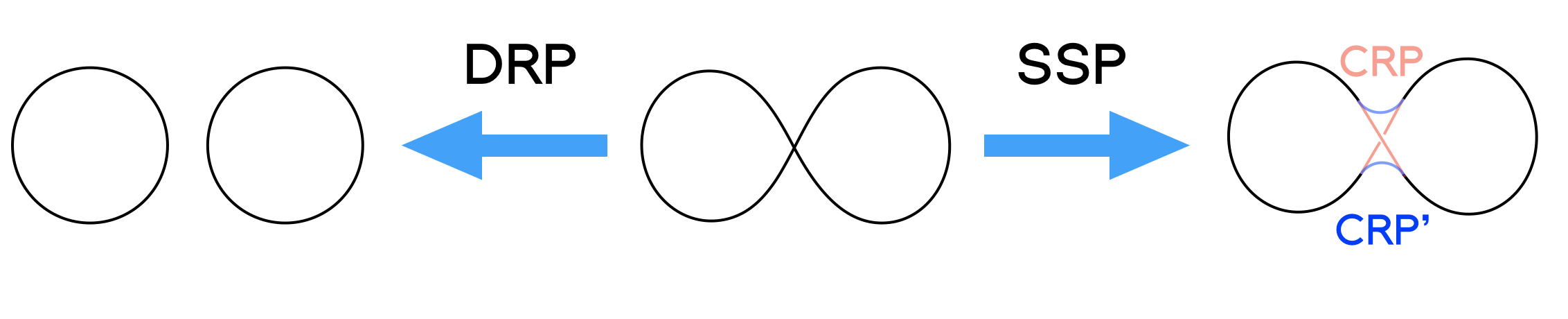}
\caption{\small Resolutions of $\S^1\vee\S^1$ and the corresponding DRP (disconnected resolution property) and SSP (strong smoothness property), the latter being a combination of two CRPs (connected resolution properties).}
\label{fig:drp-ssp}
\end{center}
\end{figure}

A key ingredient, which defines the compactification on $\8$, is that different M-theory fields satisfy different types of boundary conditions associated with the possible resolutions of the joining point; see Figure \ref{fig:drp-ssp}. The SSP (for strong smoothness property) fields propagate over a single $\S^1$ arising from the connected resolution of the two $\S^1$'s in either of the two possible combinations of orientations, the CRP and CRP$'$ (for connected resolution property), which are related by an orientation flip in one of the two $\S^1$'s, while remaining periodic with respect to the individual $\S^1$'s. On the other hand, fields obeying the DRP (disconnected resolution property) boundary condition live on either of the two disconnected $\S^1$'s upon resolving the joining point.

\begin{figure}[htb]
\begin{center}
\includegraphics[scale=.35]{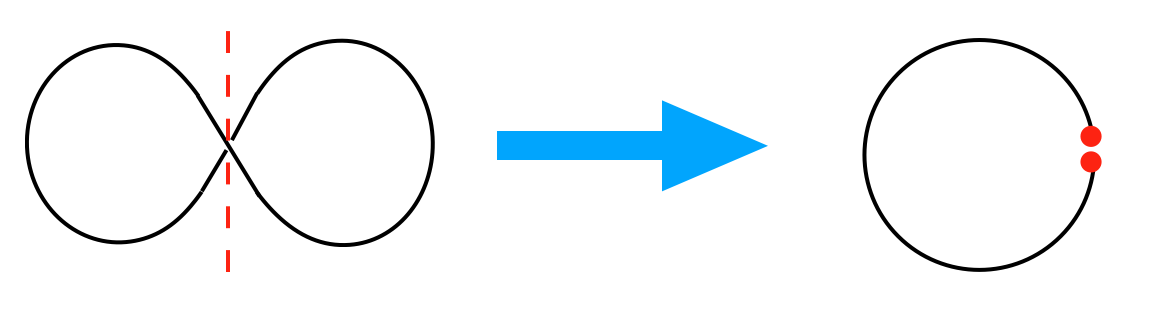}
\caption{\small The quotient of $\8$ by the $\Z_2$ exchanging the two $\S^1$'s results in an interval curled up onto itself, with its two boundaries being identified.}
\label{fig:s1vs1-z2-exchange}
\end{center}
\end{figure}

In \cite{Baykara:2026gem,Altavista:2026evd,Baykara:2026vdc}, several other maps to string theories were proposed to arise from $\Z_2$ quotients. One of these, in \cite{Baykara:2026vdc}, identifies the 10d E-type heterotic strings with the quotient of M-theory on $\8$ by the $\Z_2$ symmetry exchanging the two $\S^1$'s. The key idea is that one can picture $\8$ as a single circle (the recombined one) with two points identified. Then, in order to perform the $\Z_2$ quotient, one can first take the quotient in the recombined $\S^1$ and then identify the two points. This leads to a picture in which there is a single $\S^1$ minus a point, namely an interval curled up onto itself, with two boundaries that are forced to be identified and yet retain the memory of their distinctness. This is outlined in Figure \ref{fig:s1vs1-z2-exchange}.

The configuration is, in a sense, a quantum version of the Ho\v{r}ava--Witten theory, in which the interval is curled up onto itself, and the two boundaries are stuck together in the sense mentioned above. In~\cite{Baykara:2026vdc}, it was argued that the different 10d heterotic strings of E-type arise from choices of gauge sectors assumed to result from this configuration of coincident boundaries. A remarkable feature emerges from this setup, in particular from the requirement that there must be fields transforming under gauge factors from different boundaries: there is some notion of locality in the configuration that allows nearly coincident (yet separated) points---in this case, the boundaries---to yield light fields. 

The description of the D-type heterotics follows from the analogy with the cases of the $Spin(32)/\Z_2$ and $E_8\times E_8$ theories. Upon compactification on an $\S^1$ with suitable Wilson lines, the $E_8\times E_8$ turns into an $SO(16)\times SO(16)$ type I' configuration, with two stacks of 16 D8-branes on top of the O8-planes at each side of the interval. When the interval shrinks, additional winding modes become light and enhance the gauge symmetry to $SO(32)$ in the decompactification limit to the T-dual type I theory, which is S-dual to the $Spin(32)/\Z_2$ heterotic. In fact, as argued in \cite{Baykara:2026vdc}, for each E-type heterotic theory, it is possible to turn on some Wilson lines that break its gauge group to a subgroup of the gauge group of a D-type heterotic. Hence, the different D-type heterotics are expected to arise from the different E-type heterotics. However, the proposal includes neither a precise description of the required Wilson lines nor the additional winding states enhancing the gauge group to the D-type heterotic.

While the emerging picture is very appealing, it remains unsatisfactory in two aspects. First, it lacks an actual rationale for the pattern of possible gauge groups and light spectra of the different theories; hence, the explanation of the pattern of non-supersymmetric heterotic theories is incomplete. A second aspect is that the notion of identified boundaries is highly singular and demands a more detailed definition; a similar comment applies to the limit in which the D-type heterotics are claimed to arise. 

Our results show that these two aspects are related: we will provide a more appropriate definition of the configuration of stuck boundaries by explaining the possible resolutions of the singular configurations and their limits, and we will show that they can explain the pattern and properties (gauge group, tachyons, and massless fermion spectrum) of the non-supersymmetric heterotic theories.

\section{Quantum geometry of E-type heterotics}
\label{sec:e-limit}

We now develop our strategy to analyze the quantum-geometric realization of the E-type heterotics and to provide a microscopic interpretation of the resulting gauge groups and spectra of tachyonic and massless fields. 

\subsection{The general strategy}
\label{sec:e-strategy}

Part of the difficulty in providing a microscopic picture of the configurations reviewed in section \ref{sec:review} is that singular configurations are generally difficult to understand in M-theory. This motivates the use of standard dualities in a similar configuration in 10d type IIA string theory. This will prove extremely useful for formulating the key ingredients necessary to understand the pattern of different heterotics from M-theory on the $\Z_2$ quotient of $\8$. We proceed in several steps, which we will describe in detail now.

\subsubsection{From M-theory to the IIA ouroboros}
\label{sec:ouroboros}

In order to address the microscopic formulation of the configuration in a more tractable framework, we phrase it in terms of 10d type IIA string theory. Explicitly, we compactify the M-theory configurations on a further $\S^1$ and then shrink it to achieve a 10d type IIA setup. The resulting configuration is 10d type IIA on an interval curled up onto itself with two boundaries (defined by O8-planes, possibly with D8-branes on top) being identified. We refer to this configuration as the (IIA) ouroboros for short. As we will explain, the IIA ouroboros configuration encodes very efficiently the information about the M-theory configuration, which is unveiled by taking suitable strong coupling limits. Since these limits lead to the realization of the different 10d E-type non-supersymmetric string theories, we refer to them as E-limits.

\subsubsection{Quantum geometry and the different ouroboric variants}
\label{sec:variants}

The configuration of M-theory on an interval curled up onto itself with its boundaries stuck together, considered in \cite{Baykara:2026vdc}, is singular; therefore, the naive IIA ouroboros configurations are also singular. In particular, the endpoints of the interval play a dual role: they define the boundaries of the interval, and at the same time they define the identification points of the underlying $\8$. One may think that this is natural from the perspective of the quotient of $\8$ by the $\Z_2$ exchanging the two $\S^1$'s reviewed in section \ref{sec:review}. However, we emphasize that, from the perspective of the quotient theory, it is more natural to regard this configuration as a limit of a more general one in which the boundaries and the identification points are very close but not exactly coincident. We will show that considering different such resolutions is the key to understanding the different E-type heterotic theories.

\begin{figure}[htb]
\begin{center}
\includegraphics[scale=.31]{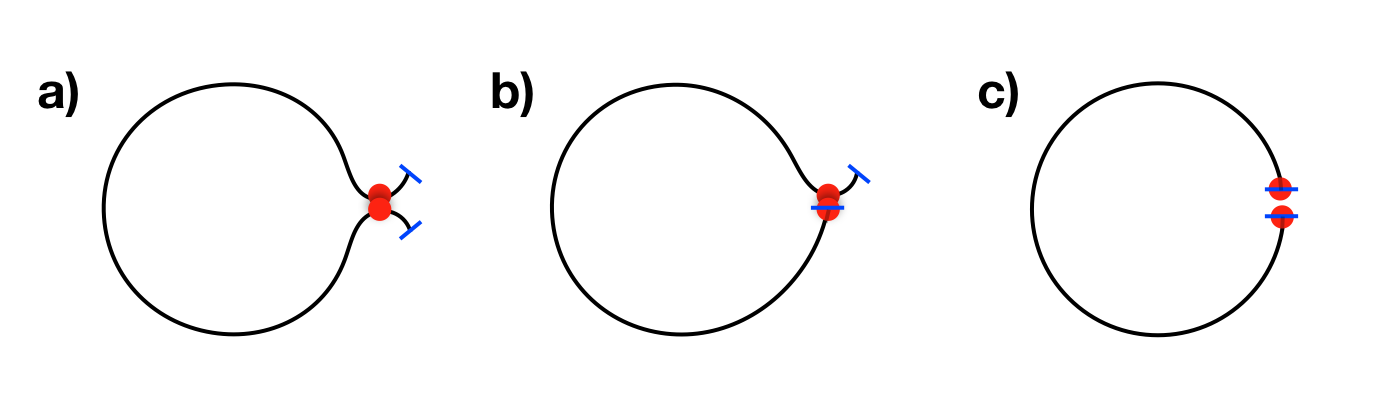}
\caption{\small The ouroboric variants. In (a), the two identification points are slightly away from the boundary points; in (b), one of the identification points is exactly one of the boundary points, even before the identification limit; in (c), both identification points are exactly the boundary points. The blue short lines correspond to boundary points and the red dots correspond to identification points.}
\label{fig:ouroboric-variants}
\end{center}
\end{figure}

Let us consider the possible configurations that correspond to an interval curled onto itself, with identification points close to (but not necessarily on top of) the boundaries. We refer to these resolutions of the ouroboros as ouroboric variants, which produce the featureless ouroboros in Figure~\ref{fig:s1vs1-z2-exchange} in the limit in which the boundaries and identification points coincide. The various ouroboric variants are shown in Figure~\ref{fig:ouroboric-variants}.

\begin{figure}[htb]
\begin{center}
\includegraphics[scale=.4]{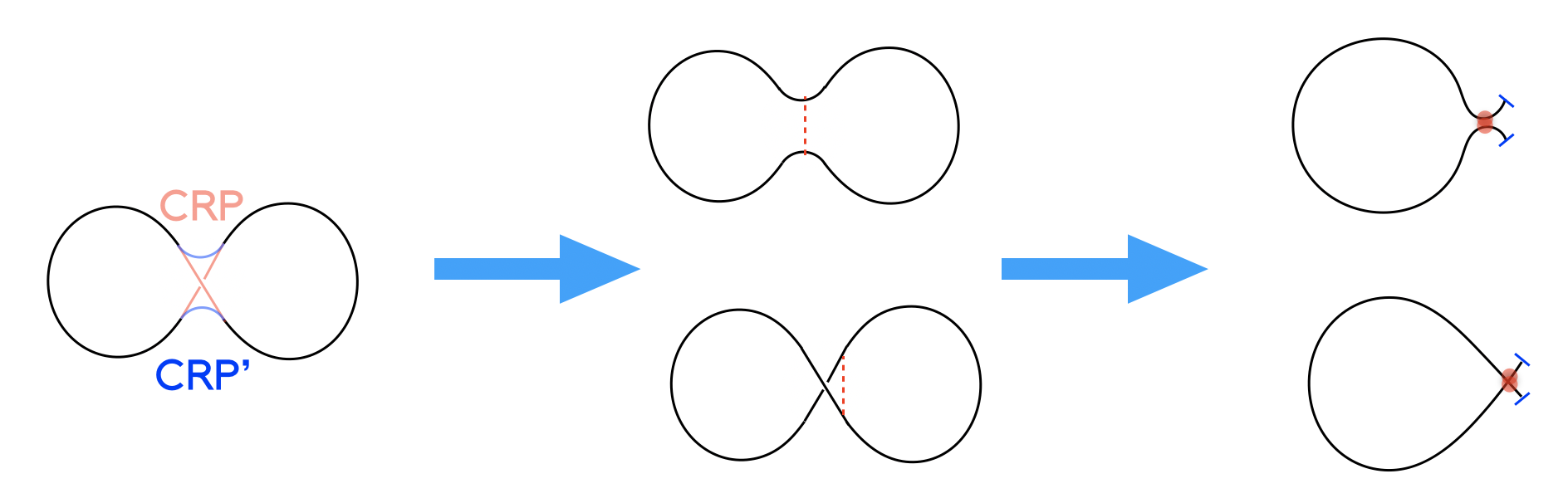}
\caption{\small The two possible quotients after exchanging the circles in $\8$ in the CRP and CRP' connected resolutions. The two resulting ouroboric variants are related by the exchange of the boundary points, but not of the identification points.}
\label{fig:quantum-z2}
\end{center}
\end{figure}

There is an additional important feature of the above ouroboric variants. The interval with the two glued boundaries arises when one quotients $\8$ by the exchange of the two circles in the connected resolution, as suggested in Figure \ref{fig:s1vs1-z2-exchange}. However, the connected resolution of $\8$ is actually a superposition of two different CRP and CRP' resolutions; see Figure~\ref{fig:drp-ssp}. Therefore, the quotient by the exchange of the circles should be regarded as a superposition of the two possible quotients. In the simplified picture of the quotient, in which boundaries and identification points are the same, there is no difference. But in our more detailed picture, in which identification points and boundaries may not be the same, the two quotients lead to two different configurations, as shown in Figure~\ref{fig:quantum-z2}.

Note that this difference arises only in ouroboric variants in which the identification points are different from the boundary points (and, in a suitable sense, also when they are both coincident), but not when one boundary coincides with one identification point while the other does not---the two configurations would be different. In other words, the superposition of the different resolutions applies to $\Z_2$-symmetric ouroboric variants. We will argue that the superposition of these two different possibilities implements a $\Z_2$ symmetry (which we dub quantum superposition $\Z_2$), which quotients the theory by an (inner or outer) automorphism of the gauge group, and hence is key to determining the global form of the gauge group.

This quantum superposition $\Z_2$ is different from the `classical' $\Z_2$ reflection symmetry of theories with two identical gauge sectors, $G\times G$, which extends the gauge group to the semidirect product $G\times G\rtimes \Z_2$. We dub it `classical' because this $\Z_2$ is already present in geometric configurations, such as the Ho\v{r}ava--Witten theory, and underlies the global structure of the supersymmetric heterotic $E_8\times E_8\rtimes \Z_2$ string. In contrast, the new quantum $\Z_2$ arises only in ouroboros compactifications (albeit with symmetric boundaries).

\subsubsection{The local picture}
\label{sec:local}

\begin{figure}[htb]
\begin{center}
\includegraphics[scale=.2]{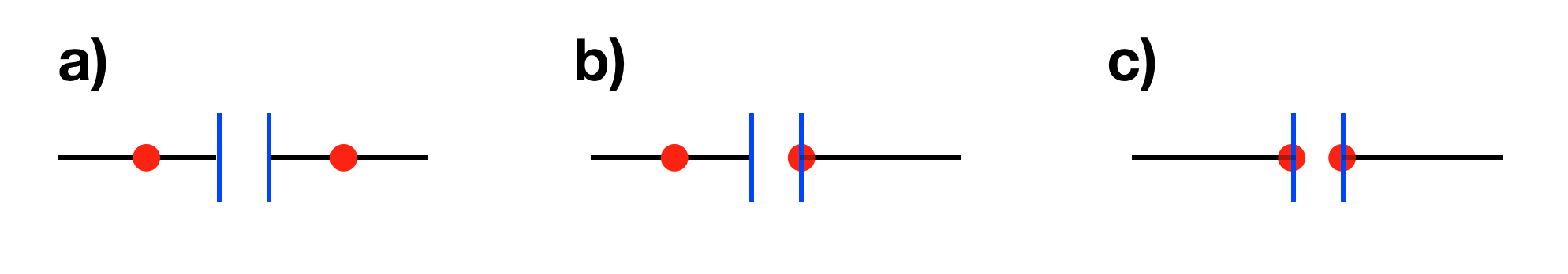}
\caption{\small Capacitor configurations resulting after zooming into the region near the boundaries (blue lines) and identification points (red dots). The labels (a), (b), and (c) agree with the corresponding ones in Figure \ref{fig:ouroboric-variants}.}
\label{fig:capacitors}
\end{center}
\end{figure}

As already pointed out in our review in section~\ref{sec:review}, in the context of the configuration in~\cite{Baykara:2026vdc}, there exists some notion of locality in the geometry of the configuration. In other words, it is reasonable to expect that the relevant physical properties of the model are determined by the structure of the configuration near the endpoints of the ouroboros, namely the pattern of boundaries and identification points in the ouroboric variants. Therefore, we may zoom into that region and focus on the resulting configurations, which we dub capacitor diagrams, shown in Figure~\ref{fig:capacitors}.

It is interesting to note that we end up with a configuration similar to two disconnected non-compact type I' theories (save for the connection via the identification points). This suggests that the interpretation of the configuration will benefit from intuition from type I' string theory; in other words, from type IIA in the presence of O8-planes and D8-branes (O8s and D8s in what follows). However, the configuration retains enough information about its non-geometric origin, as we will specify later on.

\begin{figure}[htb]
\begin{center}
\includegraphics[scale=.35]{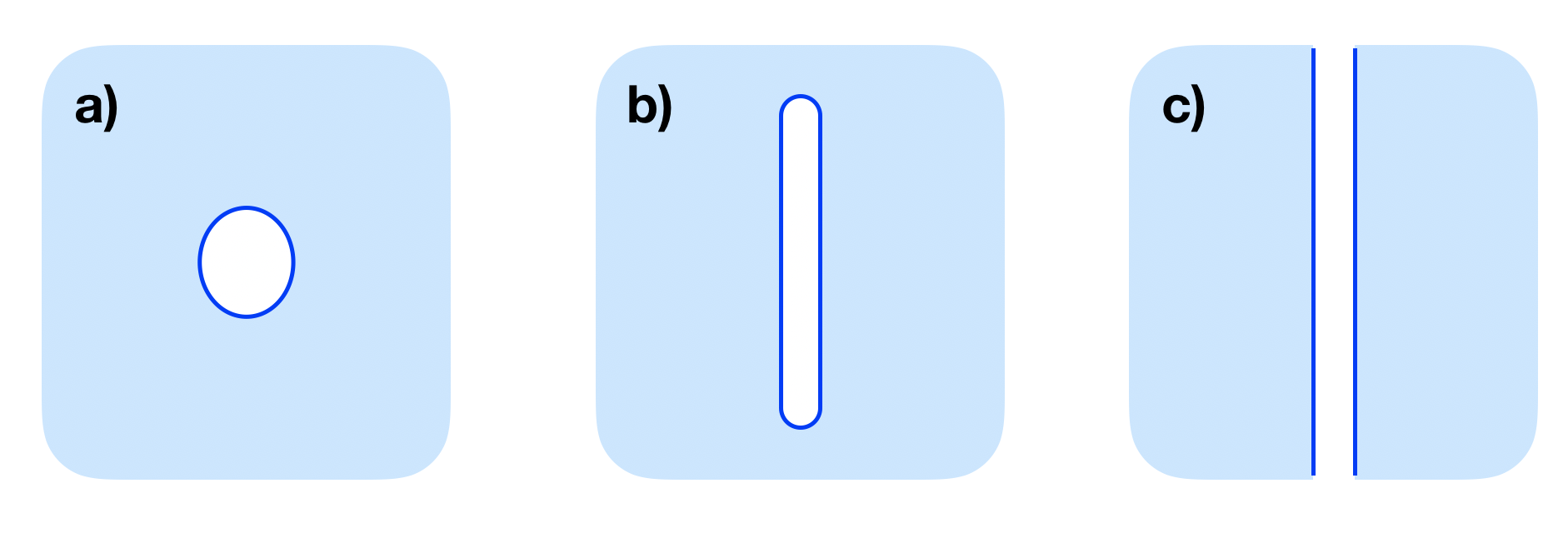}
\caption{\small Heuristic picture of the capacitor configuration as a limit of the nucleation of an infinitely elongated bubble of nothing.}
\label{fig:bubble-capacitor}
\end{center}
\end{figure}

The capacitor diagram, regarded as two disconnected type I' configurations or equivalently as a 10d type IIA theory with a thin gap splitting it into two bounded regions, motivates a suggestive heuristic picture for the interpretation of this gap. Consider type IIA in flat 10d spacetime in which we nucleate a bubble of nothing with an extremely elongated shape. Locally, it seems closely related to a capacitor configuration (see Figure \ref{fig:bubble-capacitor}), which can thus be regarded as two back-to-back end-of-the-world boundaries. 

This strongly suggests that one should regard the two sides of the capacitor configuration as oppositely oriented. In other words, the objects (O8 and D8s) located on one side should be regarded as the antiobjects (anti-O8 and anti-D8s) located on the other. More precisely, what the picture suggests is that the two sides correspond to objects and antiobjects in terms of the physics of the local capacitor configuration, but this does not establish that the same must hold for the whole (quantum) geometry in the global ouroboros. Therefore, we interpret the configuration by considering the gap between the two sides of the capacitor to be separated by a branch cut, such that the degrees of freedom arising from both sides across the gap perceive them as object/antiobject pairs. Although this does not modify the local picture, it does make a difference in the description of the global ouroboros geometry.

We envision this branch cut as being related to the odd spin structure obeyed by the SSP fermion fields, which propagate in the resolved $\S^1$ of the underlying $\8$.
Namely, in the ouroboros configuration, this odd spin structure may be regarded as being concentrated in the gap region. The appearance of objects and antiobjects would then be analogous to the object-antiobject nature of the boundaries of the Fabinger--Ho\v{r}ava version of M-theory on $\S^1/\Z_2$ with odd spin structure~\cite{Fabinger:2000jd} and, more generally, of the `M-theory breaking' of~\cite{Antoniadis:1998ki}, albeit with a quantum version of an interval curled up onto itself and with stuck boundaries.

The presence of this object-antiobject configuration across the gap explains why the local configuration is non-supersymmetric. It also implies that, as we will argue later on, the configurations generically contain tachyons, which are essentially associated with instabilities against annihilating the boundaries (or part of the degrees of freedom located therein). Interestingly, the capacitor picture also suggests that each of its sides may enjoy some remnant of supersymmetry. Indeed, this will be an important heuristic guide in our description of the appearance of the heterotic spectra.

\subsection{The microscopic picture}
\label{sec:e-microscopic}

We now address the question of the microscopic picture of the different capacitor configurations and how they describe the E-type heterotics.

\subsubsection{Gauge symmetry and enhancement at strong coupling}
\label{sec:seiberg-e-limit}

In our arguments, we use an important intuition borrowed from the supersymmetric type I' theory, in particular related to the behavior of O8/D8 systems at strong coupling and exceptional gauge symmetries, which we review in appendix~\ref{app:enhancement} following~\cite{Seiberg:1996bd} (see~\cite{Bergman:1997py,Bachas:1997kn} and \cite{Morrison:1996xf,Intriligator:1997pq,Aharony:1997ju,Aharony:1997bh} for related results in other approaches). 

In a configuration with an O8 and 16 D8s, the dilaton is constant only if the 16 D8s are exactly on top of the O8. When the D8s move off the O8, they source the dilaton, whose profile varies in a piecewise linear fashion. It is then possible to tune the D8 positions and the asymptotic dilaton (for example, on the other orientifold) to make the local value of the coupling become large at the O8 location. This leads to an enhancement of the gauge symmetry, from a generic $SO(2n)\times U(1)^k$ symmetry (for $2n$ D8s on top of the O8 and $2k$ D8s ($k$ and their $k$ orientifold images) away from it) to $E_{n+1}$ ($\times U(1)^{k-1}$). With a large asymptotic dilaton in the limit, the configuration uplifts to 11d M-theory with a Ho\v{r}ava--Witten wall, whose $E_8$ gauge symmetry is partially broken by Wilson lines of the compactification. In other words, the oft-used image that the Ho\v{r}ava--Witten wall is achieved by locating 16 D8s at the O8 and cranking up the string coupling is a simplification, and a more elaborate process is involved in the limit.

For example, as explained in appendix~\ref{app:enhancement}, the proper way to generate an $E_8$ symmetry from an O8/D8 system is to consider an O8 with 14 D8s on top and 2 D8s (one plus one orientifold image) away from it. The generic gauge symmetry is $SO(14)\times U(1)$, but it enhances to $E_8$ in a tuned limit in which the 2 D8s approach the O8 while the asymptotic dilaton grows so as to make the effective local coupling at the O8 location blow up. A similar story applies when realizing $E_7\times SU(2)$ by starting with $SO(12)\times U(1)\times U(1)$ (where the last two factors enhance to $SU(2)\times U(1)$ for coincident D8s) and taking a similarly tuned limit.

We will adopt a similar perspective to explain the gauge symmetries in our non-supersymmetric O8/D8 capacitor configurations in the local version of the ouroboros and to reproduce the gauge symmetries and light matter content of the E-type heterotic strings in an analogous tuned strong coupling limit (which we dub the E-limit).

Note that the applicability of the results of the geometric and supersymmetric case to the non-geometric and non-supersymmetric one is far from obvious. In particular, we will be forced to introduce some ad hoc assumptions about, e.g., the spectrum and behavior of bosons vs fermions in our non-geometric setup. Therefore, the reader is invited to consider the extension of standard intuitions to the non-geometric setup as a suggestive working hypothesis. However, we will find that with a remarkably simple (and physically reasonable) set of rules to construct the different capacitor configurations and to read out their gauge symmetries and matter content in the E-limit, the whole pattern of gauge symmetries and light matter content of the E-type heterotics arises in a strikingly simple way. Ultimately, this points to a more fundamental version of the rules and procedures in defining the behavior of M-theory on quantum geometries with boundaries. Their full explanation should await a better microscopic understanding of M-theory.

\subsubsection{The rules}
\label{sec:e-rules}

As we mentioned, the different E-type heterotic strings arise from a simple set of capacitor configurations of O8/D8s, together with boundaries and identification points, using a set of rules that we explain in this section. Even though they may look involved, applying them to derive the heterotic theories (in section~\ref{sec:e-heterotics}) is fairly simple.

Let us start by describing the configurations. The capacitor configurations contain O8s at the two boundaries, possibly with some (not necessarily equal) number of D8s. In addition, there are two special points (one per side of the capacitor) that are identified (hence dubbed identification points) and that may or may not coincide with the boundary points. 

Finally, we must specify the location of the D8s away from the O8. Such D8s must approach the O8 in the E-limit to recover the E-type heterotic theories. This is reminiscent of the fact that the identification points must also approach the boundary (i.e., the O8) in the limit in which we recover the ouroboros configuration in \cite{Baykara:2026vdc}; namely, the quotient of $\8$ by the exchange of the $\S^1$'s. The simplest assumption is that these two limits are the same, and therefore that the identification points and the D8s away from the O8 are located at the same position.

Let us now fix some nomenclature: we refer to boundaries that coincide with an identification point as \emph{glued}, and boundaries that do not coincide with an identification point as \emph{detached}. We also refer to identification points that are away from the O8 as \emph{unorientifolded}, while those on top of an O8 are \emph{orientifolded} and correspond to a glued O8. Finally, recall that the O8/D8s on the two sides of the local capacitor picture should be regarded as objects and antiobjects; more precisely, this comes from the presence of a branch cut in the gap that separates them, enforcing the fact that degrees of freedom connected to both sides perceive them as objects and antiobjects. With this understanding, we will refer to all objects as O8/D8s, hoping that this leads to no confusion.

We now present the rules with some heuristic motivation for them: 

\begin{enumerate}

\item Coincident D8-branes in a given stack lead to gauge sectors, but do not give rise to fermions or scalars (including tachyons), which arise from other sectors to be specified later on. Explicitly, $2n$ D8s on top of the O8 lead to $SO(2n)$ gauge symmetry, while $2k$ D8s ($k$ and their $k$ orientifold images) at an unorientifolded identification point (i.e., away from the O8) lead to a $U(k)$ gauge symmetry. The $SO(2n)\times U(1)$ symmetry (with the $U(1)$ from $U(k)\simeq SU(k)\times U(1)$) is enhanced to $E_{n+1}$ in the strong coupling E-limit. These rules are inspired by the usual supersymmetric type I' context (except that we only get gauge bosons, and no adjoint matter fields).
\label{e-rule-coincident-branes}

\item There is no matter between the D8s in two detached boundaries (i.e., away from identification points). This rule is sensible because boundaries are separated by the gap, and if they are detached, they cannot coincide through the identification of points.
\label{e-rule-detached-boundaries}

\item Between the D8s in two glued boundaries (i.e., on top of identification points), we obtain (negative chirality\footnote{Even though we are in 9d, the chirality of the rules in this section should be regarded as well-defined chiralities in the 10d theory in the E-limit, which match those of the corresponding E-type heterotics.}) fermions in the bifundamental of the $SO\times SO$ symmetry. This rule is similar to the appearance of bifundamental fermions in coincident brane-antibrane pairs. We interpret the fact that there are, however, no tachyons as the statement that the brane-antibrane pair is `protected' by the presence of both O8s.
\label{e-rule-glued-boundaries} 

\item Between the D8s on top of identification points, we obtain tachyons in the bifundamental unless both are orientifolded identification points (i.e., glued boundaries; see rule \ref{e-rule-glued-boundaries}). 
\label{e-rule-tachyon}

The above two rules, \ref{e-rule-glued-boundaries} and \ref{e-rule-tachyon}, specify the spectrum of branes and antibranes at identification points. If they are both orientifolded, the tachyon is removed, and we obtain bifundamental fermions. If at least one of them is unorientifolded, the protection disappears and we get a bifundamental tachyon.

\item Between the D8s on a detached boundary and the D8s at unorientifolded identification points (on the same side or on the other side), we get fermions transforming under the gauge sector of the boundary and the gauge sector of the D8s at the identification points (in all possible bifundamental representations). 
This rule is motivated as an extension of rule \ref{e-rule-glued-boundaries}: in the limit, the D8s at the unorientifolded identification points approach the detached boundary. 

The fermions have positive chirality if both gauge factors are on the same side of the capacitor diagram and negative if they are on different sides. This assignment agrees with the intuition that degrees of freedom crossing the gap feel the D8s as brane-antibrane pairs, i.e., have flipped chiralities as compared to the brane-brane case.
\label{e-rule-detached-boundary-unorientifolded-point}

\item Finally, there are stuck D0-branes on top of each O8. If the O8 is a detached boundary (i.e., not an identification point), the D0s have fermion zero modes under the D8s located on top of the O8, and upon quantization, they provide the spinor states in the enhancement to the exceptional gauge symmetry in the E-limit. 

In fact, since the D8s at the unorientifolded identification point(s) approach the O8 in the E-limit, D0-branes also have zero modes charged under them. Upon their quantization, the spectrum of D0-brane states yields massless fermions charged in a simultaneous spinor representation of the different gauge factors, which partner with the fermions from rule \ref{e-rule-detached-boundary-unorientifolded-point} to complete the representations of the enhanced exceptional symmetry in the E-limit.

There is an overall $\Z_2$ projection correlating the spacetime chirality of the fermion and the (overall) chirality of the spinor representation under the gauge groups.
\label{e-rule-D0s-detached}

\item At a glued boundary, the D0-brane has fermion zero modes charged under the D8s (those at the glued boundary itself, and also those at the other unorientifolded identification point); upon quantization, they provide fermion matter states in the spinor representations (with chiralities determined as in rule \ref{e-rule-D0s-detached}).
\label{e-rule-D0s-glued}

\item In symmetric capacitor configurations, one should account for the quantum superposition $\Z_2$, arising from the CRP and CRP' resolutions of the underlying $\8$, as a quotient by an (inner or outer) automorphism of the gauge group; see section~\ref{sec:variants}. This can determine the global structure of the gauge group, or can lead to the combination of different gauge factors.
\label{e-rule-crps}

\end{enumerate}

We should note that, although the rules above are motivated by analogies with rules governing open strings and D0-brane states in conventional setups, we are extending their application to non-geometric (and non-supersymmetric) configurations. This implies important modifications; for instance, for coincident identical branes in rule \ref{e-rule-coincident-branes}, we obtain gauge bosons but require that there are no adjoint fermions or scalars. Moreover, in the rules specifying the fermion content from different sectors, we require that they do not produce massless scalars; similarly, in rule \ref{e-rule-tachyon}, the sector producing tachyons does not lead to any fermions. We admittedly do not have a full explanation for these facts, but we emphasize that this very different behavior of sectors of gauge bosons, scalars, and fermions is to be expected in a non-geometric context. Actually, we find it remarkable that the rules producing the right pattern of non-supersymmetric E-type heterotic strings are so reminiscent of the rules in conventional setups.

\begin{figure}[htb]
\begin{center}
\includegraphics[scale=.4]{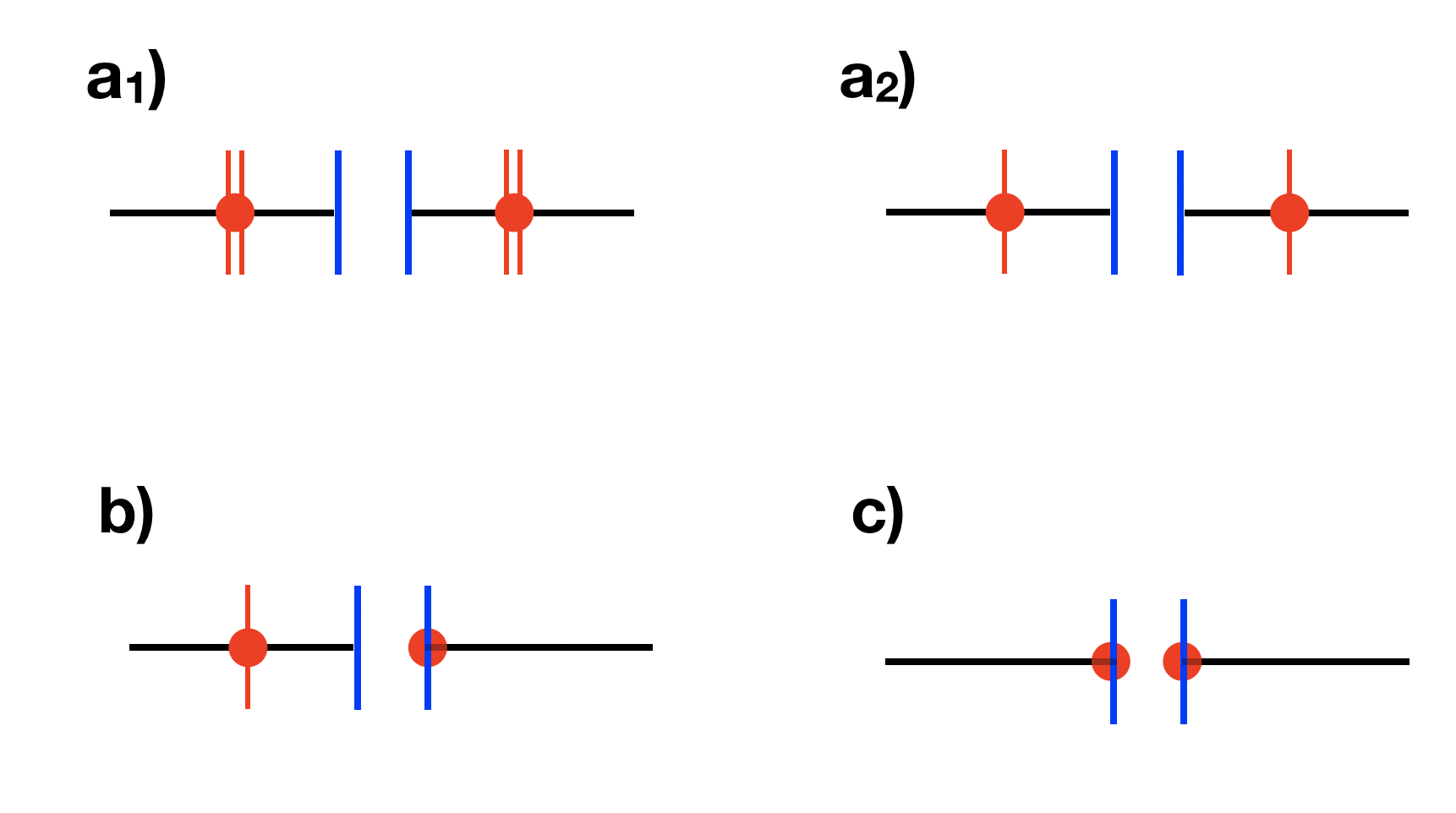}
\caption{\small Dressed capacitor configurations obtained by specifying the D8 locations (red lines) at \emph{unorientifolded} identification points in the capacitor diagrams of Figure \ref{fig:capacitors}. We again indicate boundaries with blue lines and identification points with red dots. When a blue line and a red dot coincide, they are called glued boundary and orientifolded identification point, while if they don't, they are called detached boundary and unorientifolded identification point. The labels (a), (b), and (c) agree with those in Figures \ref{fig:capacitors}, \ref{fig:ouroboric-variants}, with the case (a) admitting two variants. For clarity, we have not indicated the numbers of D8s at the boundaries (they are fixed by the total of 16 in the covering space).}
\label{fig:dressed-capacitors}
\end{center}
\end{figure}

Finally, note that the configurations leading to E-type heterotic strings correspond to a very restricted set of possible D8-brane distributions. In fact, the number of D8-branes located at an unorientifolded identification point is at most 4 ($2+2$ orientifold images) if the other identification point is unorientifolded as well (Figures \ref{fig:ouroboric-variants}a or \ref{fig:capacitors}a), whereas it is fixed at 2 ($1+1$ orientifold image) if the other identification point is a glued boundary (Figures \ref{fig:ouroboric-variants}b or \ref{fig:capacitors}b). The resulting four possible configurations are shown in Figure \ref{fig:dressed-capacitors}.

This constraint ensures that we recover only the E-type heterotics and no other exotic theories. It would be interesting to gain a better understanding of the origin of this restriction. The simplest explanation would be that other distributions of D8-branes are allowed, but they do not lead to new E-limits beyond those that we have determined. This would be analogous to the fact that in the geometric setup, different distributions of D8s near an O8 can lead to different enhanced exceptional symmetries, yet they can all be interpreted as the breaking of a unique $E_8\times E_8$ theory by Wilson lines. 
An alternative possibility is that D8 configurations in the quantum geometry satisfy some sort of exclusion principle, producing a bound in the number of D8s located at unorientifolded identification points. This would be analogous to the bounds on D-brane charges in WZW models; see, e.g., \cite{Maldacena:2001xj} and references therein. It would be interesting to gain further insight into these (or other) possible explanations for the restricted set of heterotic theories which can be built from this quantum-geometric perspective. 

\subsection{E-type heterotics}
\label{sec:e-heterotics}

We now apply the rules from the previous section to obtain the spectra of the different E-type heterotic theories. 

\subsubsection{The \texorpdfstring{$E_8\times SO(16)$}{E8 x SO(16)} theory}
\label{sec:e8so16}

We start with the $E_8\times SO(16)$ theory and show that it arises from an ouroboros of type (b) in Figure \ref{fig:ouroboric-variants}, namely a capacitor of type (b) in Figures \ref{fig:capacitors}, \ref{fig:dressed-capacitors}. There is one detached boundary containing an O8 and 14 D8s, and there are 2 D8s (one and its orientifold image) at the unorientifolded identification point, which is glued to the second boundary that contains an O8-plane and 16 D8s.

The computation of the spectrum is as follows. Following rule \ref{e-rule-coincident-branes}, there is a gauge symmetry $SO(14)\times U(1)$ on the side with the detached boundary and an $SO(16)$ on the side of the glued boundary. In the detached boundary, the symmetry enhances to $E_8$ in the E-limit, with D0-branes providing the spinorial gauge bosons according to rule \ref{e-rule-D0s-detached}. Hence, the total symmetry is $E_8\times SO(16)$. 

Regarding matter, rule \ref{e-rule-tachyon} implies that there is a tachyon in the bifundamental of the $U(1)$ and the $SO(16)$ gauge symmetries at the identification points; namely, we have a tachyon in the ${\bf 16}$ of $SO(16)$. In this case, there are no fermions arising from rule \ref{e-rule-detached-boundary-unorientifolded-point}, but there are fermions arising from the D0-branes on top of the glued boundary according to rule~\ref{e-rule-D0s-glued}. Specifically, the D0-brane has fermion zero modes charged under $SO(16)$ and $U(1)$ at the unorientifolded identification point. Their quantization leads to fermions in the ${\bf 128}+{\bf 128'}$ of $SO(16)$, with correlated spacetime and $SO(16)$ chiralities. This agrees with the spectrum in Table \ref{table:heterotics}.

In this case, the geometry is not affected by the quantum superposition $\Z_2$  in rule \ref{e-rule-crps}, so the global structure of the gauge group is simply $E_8\times Spin(16)$, which is consistent with~\eqref{global}.

\subsubsection{The \texorpdfstring{$[E_7\times SU(2)]^2$}{[E7 x SU(2)]2} theory}
\label{sec:e7su2e7su2}

We now consider the $[E_7\times SU(2)]^2$ theory, which, as we will show, arises from an ouroboros of type (a) in Figure \ref{fig:ouroboric-variants}, i.e., a capacitor of type (a) in Figure \ref{fig:capacitors}, and in particular of type (a1) in Figure \ref{fig:dressed-capacitors}.
The two boundaries are detached, and each contains an O8 with 12 D8-branes on top. There are 4 D8-branes (2 and their 2 orientifold images) at each unorientifolded identification point.

The computation of the spectrum is as follows. Using rule \ref{e-rule-coincident-branes}, there is a gauge symmetry $SO(12)\times U(1)\times SU(2)$ on each side of the capacitor. Since the boundaries are detached, the symmetry enhances to $[E_7\times SU(2)]^2$ in the strong coupling E-limit (with D0-branes providing the spinorial gauge bosons, according to rule \ref{e-rule-D0s-detached}). 

As for the matter content, rule \ref{e-rule-tachyon} implies that there is a tachyon in the bifundamental of the $SU(2)^2$. Finally, using rule \ref{e-rule-detached-boundary-unorientifolded-point}, we obtain fermions in the bifundamentals of the $SO(12)$'s and the $SU(2)$'s (with positive or negative chirality for `same-side' or `different-side' gauge factors). Using rule \ref{e-rule-D0s-detached}, we find a similar pattern of charged fermions as doublets under the $SU(2)$'s but in the spinor representation ${\bf 32}$ of the $SO(12)$ (with the spacetime chirality correlated with the $SU(2)^2=SO(4)$ chirality). Upon enhancement to $E_7$ in the E-limit, these fields combine into fermions in the $({\bf 56},{\bf 2};{\bf 1},{\bf 1})_++({\bf 56},{\bf 1};{\bf 1},{\bf 2})_-$, and in the $({\bf 1},{\bf 2};{\bf 56},{\bf 1})_-+({\bf 1},{\bf 1};{\bf 56},{\bf 2})_+$ (with subindices indicating spacetime chirality). The complete spectrum agrees with that of the $[E_7\times SU(2)]^2$ theory in Table \ref{table:heterotics}.

We should now take into account that the geometry has two detached boundaries; therefore, we must include the effect of the quantum superposition $\Z_2$ in rule \ref{e-rule-crps}. As explained, this should amount to quotienting the group by an automorphism of the gauge group, which we now argue determines part of the global structure of the gauge group. 
The gauge factors in this model have a non-trivial center, namely $Z(SU(2))=\Z_2$---with doublets being odd and triplets being even---and $Z(E_7)=\Z_2$---with the ${\bf 56}$ being odd and the adjoint being even.\footnote{Actually, focusing on a single side of the capacitor diagram, hence considering states that live entirely on that side, the presence of fermions in the $({\bf 56},{\bf 2})$ of the $E_7\times SU(2)$ implies that the gauge group at each side is the quotient $(E_7\times SU(2))/\Z_2$ by the diagonal $\Z_2$.} We expect the quantum superposition $\Z_2$ to correspond to a suitable subgroup of these symmetries.  

Let us now consider the effect of the quantum superposition. Interestingly, in this case, the quantum superposition implies the exchange of the two boundaries, but not of the identification points, and so it morally exchanges the $E_7$'s while leaving the $SU(2)$'s untouched. In fact, this explains why the spectrum contains not only fermions in the $({\bf 56},{\bf 2};{\bf 1},{\bf 1})$, $({\bf 1},{\bf 1};{\bf 56},{\bf 2})$, i.e., charged under `same-side' gauge factors, but also fermions in the $({\bf 1},{\bf 2};{\bf 56},{\bf 1})$, $({\bf 56},{\bf 1};{\bf 1},{\bf 2})$, i.e., charged under `different-side' gauge factors. In group theory terms, this implies that the global structure of the gauge group is $[E_7\times SU(2)]^2/\Z_2$, where the $\Z_2$ is the diagonal combination of the two $\Z_2$'s on the two sides of the capacitor (equivalently, the diagonal combination of the centers of the four gauge factors). Therefore, the quantum superposition determines the global structure of the gauge group by identifying ingredients from both sides. Note that this superposition provides a further rationale for the fact that degrees of freedom on one side (such as the $SO(12)\subset E_7$ or the D0-branes responsible for the appearance of the ${\bf 56}$'s) can talk to degrees of freedom (like the $SU(2)$'s) on the other side, as is clear from the proposed rules.

The complete global structure of the gauge group, including the classical semidirect product $\Z_2$ associated with the reflection symmetry of the figure, is $[(E_7\times SU(2))^2/\Z_2]\rtimes \Z_2$, as in \eqref{global}.

It is interesting to note that the existence of a non-trivial center is already present in the theory before the E-limit. Indeed, we have $Z(Spin(12))=\Z_2\times \Z_2$ (with the elements of the group acting non-trivially on the vector, spinor, or conjugate spinor representations) and $Z(SU(2))=\Z_2$. Hence, the constraint on the global form of the gauge group, in particular the $\Z_2$ quotient relating the two sides of the capacitor picture, is already present at this level. This will be useful in the discussion of the global structure of the gauge symmetry of the corresponding D-type heterotic theory in section \ref{sec:d-heterotics} (concretely, section \ref{sec:so24so8}).

\subsubsection{The \texorpdfstring{$SO(16)\times SO(16)$}{SO(16) x SO(16)} theory}
\label{sec:so16so16-e}

We now consider the $SO(16)\times SO(16)$ theory and show that it arises from an ouroboros of type (c) in Figure \ref{fig:ouroboric-variants}, namely a capacitor of type (c) in Figures \ref{fig:capacitors}, \ref{fig:dressed-capacitors}. The two boundaries are glued, and each contains an O8-plane and 16 D8s.

The computation of the spectrum is as follows. Using rule \ref{e-rule-coincident-branes}, there is a gauge symmetry $SO(16)$ on each side of the capacitor, resulting in a total $SO(16)\times SO(16)$ gauge group, which is not enhanced. 

As for the matter content, according to rule \ref{e-rule-tachyon}, there are no tachyons between the D8s in the glued boundaries because of the presence of the O8s. On the other hand, according to rule \ref{e-rule-glued-boundaries}, there are bifundamental fermions (with negative spacetime chirality) in the $({\bf 16},{\bf 16})$ of $SO(16)\times SO(16)$. Finally, there are D0-branes on top of the O8s, with fermion zero modes charged under the corresponding D8s, according to rule \ref{e-rule-D0s-glued}. Their quantization leads to (same-chirality) fermions in spinorials $({\bf 128},{\bf 1})+({\bf 1},{\bf 128})$. This agrees with the spectrum in Table \ref{table:heterotics}.

The applicability of the quantum superposition $\Z_2$  in rule \ref{e-rule-crps} to this case of two glued boundaries would in principle be open to interpretation. On the other hand, the global gauge group in (\ref{global}) has the structure $[Spin(16)^2/\Z_2]\rtimes \Z_2$, which is very similar to that of the $[E_7\times SU(2)]^2$ theory. This suggests that, in analogy with our analysis of this theory in section \ref{sec:e7su2e7su2}, the $\Z_2$ quotient should correspond to the effect of the quantum superposition $\Z_2$.

\subsubsection{The \texorpdfstring{$(E_8)_2$}{(E8)2} theory}
\label{sec:e82}

Let us finally consider the $(E_8)_2$ theory, where the gauge factor is realized at level 2 (in the terminology of the Kac--Moody algebra in the worldsheet description). We will show that this theory arises from  an ouroboros of type (a) in Figure \ref{fig:ouroboric-variants}, i.e., a capacitor of type (a) in Figure \ref{fig:capacitors}, and more concretely (a2) in Figure \ref{fig:dressed-capacitors}.
The two boundaries are detached and contain an O8 with 14 D8-branes on top, and there are 2 D8-branes ($1+1$ orientifold image) at each unorientifolded identification point.

The computation of the spectrum is as follows. Using rule \ref{e-rule-coincident-branes}, there is a gauge symmetry $SO(14)\times U(1)$ on each side of the capacitor. Since the boundaries are detached, the symmetry enhances to $E_8\times E_8$ in the strong coupling E-limit (with D0-branes providing the spinorial gauge bosons, according to rule \ref{e-rule-D0s-detached}). This gauge group will be modified at a final stage to its diagonal $E_8$ combination, but we remain within the $E_8\times E_8$ picture for the time being.

As for the matter content, rule \ref{e-rule-tachyon} implies that there is a tachyon in the `bifundamental' of the $U(1)^2$. This is puzzling from the perspective of the enhancement of the gauge group to $E_8\times E_8$ because there are no other states to fill out the required large representations. Remarkably, the final combination of the gauge factors into a single diagonal $E_8$ solves this problem, because the `bi-fundamental' of the $U(1)$'s on different sides is a singlet. Hence, we will eventually obtain a tachyon singlet.

Using rule \ref{e-rule-detached-boundary-unorientifolded-point}, we obtain fermions in `bifundamentals' of the $SO(14)$'s and $U(1)$'s, with positive or negative spacetime chirality for `same-side' or `different-side' gauge factors. Using rule \ref{e-rule-D0s-detached}, the D0-branes also have 18 fermion zero modes charged as vectors under the $SO(14)$ and under the $U(1)$'s. Upon quantization, the D0-brane states thus provide 512 fermions transforming as simultaneous spinors of $SO(14)$ and of the $SO(2)$'s from each D8 and its orientifold image at each identification point. After the projection under the D0-brane worldvolume $O(1)\simeq \Z_2$ gauge symmetry, only 256 states survive, transforming as fermions in the ${\bf 128}_++{\bf 128}'_-$ of the $SO(16)\subset E_8$ (with subindices indicating fermion chirality). Hence, combining with the `bifundamentals' of the $SO(14)$'s and $U(1)$'s mentioned above (which are enhanced to ${\bf 120}$'s of the $SO(16)\subset E_8$), the spectrum contains opposite chirality fermions in the $({\bf 248},{\bf 1})_\pm +({\bf 1},{\bf 248})_\pm$ of $E_8\times E_8$.

We must now take into account that the geometry has two detached boundaries, and we must include the quantum superposition $\Z_2$ of rule \ref{e-rule-crps}. As in the previous example, this corresponds to identifying things via an automorphism of the gauge group. However, in contrast to previous examples, the gauge group $E_8\times E_8$ has no non-trivial center, i.e., $Z(E_8)={\bf 1}$; therefore, the quantum superposition $\Z_2$ must be acting as an outer automorphism. The effect of the quantum superposition is that the two gauge factors become identified, so that the actual gauge group is the diagonal $E_8$. As already anticipated, there is a tachyon singlet under this symmetry. Finally, the fermion matter content also experiences the quantum superposition $\Z_2$ identification and ends up transforming as 2 opposite chirality fermions in the adjoint representation. The complete spectrum agrees with that of the $(E_8)_2$ theory in Table \ref{table:heterotics}.

The projection onto the diagonal subgroup fits well with the realization of this gauge factor at level 2 from the worldsheet perspective, or equivalently, with the realization of this theory as an orbifold of the supersymmetric $E_8\times E_8$ theory by the exchange of the two factors. 

\medskip

This concludes our discussion of how the different 10d non-supersymmetric E-type heterotic theories arise from our ouroboric variant configurations, which provide a more detailed quantum resolution of the configuration in \cite{Baykara:2026vdc}. In a precise sense, we have explained the (quantum) geometry and the physics underlying the choices of boundary conditions in the proposal of~\cite{Baykara:2026vdc}, which allowed us to perform a complete characterization of the different heterotic theories, their light spectra, and the global structure of their gauge groups.

Our derivation follows the systematic application of a simple set of rules, which are physically well-motivated by their analogy with the geometric and supersymmetric setup. As a matter of fact, their implementation in a quantum-geometric and non-supersymmetric context has required the introduction of some ad hoc prescriptions regarding the sectors where gauge bosons, tachyons, and fermions arise. However, it is highly remarkable that simple (quantum) geometric prescriptions manage to produce the rich pattern of non-supersymmetric heterotic theories. We now move on to extending this understanding to the D-type heterotics.

\section{Quantum geometry of D-type heterotics}
\label{sec:d-limit}

We now turn to the discussion of D-type heterotics. As reviewed in section \ref{sec:review}, the proposal in \cite{Baykara:2026vdc} for these theories is that they are related to E-type heterotics in essentially the same way as the $Spin(32)/\Z_2$ is related to the $E_8\times E_8$. Namely, putting an E-type heterotic theory on a further $\S^1$ with suitable Wilson lines should lead to a type I'-like configuration, such that when one takes the limit in which the interval shrinks (and the coupling becomes large), the gauge group is enhanced by mixing together the two sectors of the original E-type theory, thus recovering the gauge group of some D-type heterotic theory. This is an appealing picture, but with the serious drawback that there was no microscopic description for the original E-type heterotics, hence no clear guide in the search for the suitable Wilson lines to run the argument, and no microscopic description of the degrees of freedom triggering the enhancement of the gauge group to the final D-type heterotic.

\subsection{The general strategy}
\label{sec:d-strategy}

In our approach, developed in the following, we adopt the above strategy, but with a clear advantage. We have built the E-type theories as a particular limit (the E-limit) of a type I'-like configuration, namely the ouroboric variants. This already solves one of the main ingredients that were missing in the proposal of \cite{Baykara:2026vdc}: the choice of Wilson lines. The step of compactifying the E-type heterotic on an $\S^1$ with suitable Wilson lines turned on corresponds, in our framework, to simply undoing the E-limit. Therefore, for each E-type heterotic string, we will consider the corresponding ouroboric variant configuration, and instead of taking the E-limit, we will consider a different limit (which we naturally dub D-limit) in which the length of the ouroboros shrinks. We will provide an explicit set of rules (again, inspired by the usual geometric setup) to easily read out the extra degrees of freedom that become light in the limit and that provide the enhancement of the gauge group and of the light matter content to that of a corresponding D-type heterotic theory. The D-type heterotics corresponding to the various E-type heterotics were located with hindsight in the same row in Table \ref{table:heterotics}. 

We now turn to the discussion of the rules governing the enhancement of the gauge symmetry and light matter content in the D-limits.

\subsection{The rules}
\label{sec:d-rules}

The above arguments suggest a very natural set of rules to understand the D-type heterotic theories resulting from the D-limit of the ouroboros configurations associated with a given E-type heterotic. To spell out the rules, we adopt the same terminology for boundaries and identification points as in section \ref{sec:e-rules}. A key difference is that the extra states becoming massless in the D-limit are not associated with the local capacitor diagram but rather with open string states stretched across the ouroboros circle. These degrees of freedom do not cross the gap in the capacitor diagram; therefore, as explained in section \ref{sec:local}, we propose that they perceive the ingredients on the two sides as identical objects rather than object/antiobject pairs. This is a crucial assumption to obtain the correct enhanced gauge symmetries.

We now present our rules for the physics of D-limits:

\begin{enumerate}

\item One should not consider D0-brane states, which only give massless states in the E-type strong coupling limit. They simply become massive spinorial states of the corresponding D-type heterotics.
\label{d-rule-d0s}

\item The $SO(2n_1)$ and $SO(2n_2)$ gauge factors from the two boundaries combine into a single $SO(2n_1+2n_2)$ (unless they are both glued boundaries, in which case they are `protected' against enhancement). We also prescribe that if there is an unpaired $U(1)$ (e.g., from D8s at an unorientifolded identification point glued to a boundary), it must also be included in this enhanced $SO$, thus adding one extra unit of rank. 
\label{d-rule-so} 

The two rules above are essentially as in the relation between the supersymmetric $E_8\times E_8$ and $Spin(32)/\Z_2$ heterotics via type I'; hence, they can be seen as arising from standard type I' physics (with the additional assumption that there are no scalar or fermion degrees of freedom in these enhancements in the non-geometric and non-supersymmetric context).  

\item The remaining $U(m_1)$ and $U(m_2)$ gauge factors from D8s at unorientifolded identification points (if present) combine and enhance into $SO(2m_1+2m_2)$. This can be understood as the enhancement of the symmetry of the $2m_1$ and $2m_2$ D8s (including orientifold images) as they come close to the O8s in the limit. 
\label{d-rule-u}

\item As for the matter content (either tachyons or massless fermions), their representations enhance with extra winding states to fill out a representation of the enhanced symmetry group. In particular, the number of tachyons doubles from the E-limit to the D-limit. Analogously, fermions double in the D-limit, but only after removing the spinor representations under the gauge group (see rule~\ref{d-rule-d0s}).\footnote{The chiralities of the fermions obtained in the E-limit are not preserved in the D-limit. This is not a contradiction because our setup is ultimately 9d and there is a T-duality involved in the D-limit, which corresponds to shrinking the interval, T-dualizing, and taking a strong coupling limit.} 
\label{d-rule-matter}

\end{enumerate}

As with the rules for E-limits in section \ref{sec:e-rules}, our rules for D-limits are also motivated by analogies with standard open strings. Our application to non-geometric and non-supersymmetric configurations clearly implies important modifications, such as the absence of adjoint scalars and fermions in the sector of D8s that produces gauge bosons, and the appearance of fermions in sectors with no scalars and vice versa. We admittedly do not have a full explanation for these facts, but we emphasize that it is surprising that the rules producing the right pattern of non-supersymmetric heterotic strings are so reminiscent of the rules in usual geometric setups.

In the following section, we apply the above rules to the ouroboric variants producing the E-type theories to recover the D-type heterotics in the limit. 

\subsection{D-type heterotics}
\label{sec:d-heterotics}

We now apply the rules from the previous section to obtain the different D-type heterotic theories. 

\subsubsection{The non-supersymmetric \texorpdfstring{$SO(32)$}{SO(32)} from the \texorpdfstring{$E_8\times SO(16)$}{E8 x SO(16)} theory}
\label{sec:so32}

Let us consider the application of the rules from section \ref{sec:d-rules} to the ouroboric variant (b) in Figure \ref{fig:ouroboric-variants} (see the capacitor diagrams in Figures \ref{fig:capacitors}b, \ref{fig:dressed-capacitors}b), which realizes the $E_8\times SO(16)$ theory in the E-type limit, c.f.~section \ref{sec:e8so16}. The configuration has one detached boundary with an O8 and 14 D8s and one boundary with an O8 and 16 D8s glued to an unorientifolded identification point with 2 D8s (one and its orientifold image).

The gauge symmetry of the ouroboros configuration before the E-type limit is $SO(14)\times U(1)\times SO(16)$. Using rule \ref{d-rule-so}, this is enhanced to $SO(32)$, which includes the unpaired $U(1)$. From a simplified perspective, the $E_8\times SO(16)$ has an $SO(16)\times SO(16)$ subgroup, which enhances to $SO(32)$ in the D-limit.

Let us now consider the matter content. The tachyons in the $({\bf 1}, {\bf 16})$ of the $E_8\times SO(16)$ theory descend to tachyons in the $({\bf 1}_0,{\bf 16})$ of $SO(14)\times U(1)\times SO(16)$ (with the subindex indicating the $U(1)$ charge). According to rule \ref{d-rule-matter}, their number of components should double in the D-limit, and therefore they must enhance to a tachyon in the ${\bf 32}$ of $SO(32)$. Regarding the massless fermions, all such fields in the $E_8\times SO(16)$ theory arise from D0-branes, which, according to rule \ref{d-rule-d0s}, do not contribute to massless matter in the D-type limit. Therefore, the resulting theory does not contain massless fermions. 

The D-limit has thus gauge group $SO(32)$, a tachyon in the vector representation, and no massless fermions, in agreement with the spectrum of the non-supersymmetric $SO(32)$ theory; see Table \ref{table:heterotics}. Note that, even though the D0-branes do not give rise to massless states, they provide the massive spinor states of the $SO(32)$ theory. These are stable non-supersymmetric D0-brane states in the open string theory, which therefore survive in the strong coupling limit to the D-type heterotic theory. Here, they correspond to perturbative massive spinor states, in close analogy with the stable non-BPS D0-branes of type I string theory (see \cite{Sen:1999mg} for a review). This implies that the actual global form of the gauge group is $Spin(32)$, in agreement with \eqref{global}.

\subsubsection{The \texorpdfstring{$SO(24)\times SO(8)$}{SO(24) x SO(8)} from the \texorpdfstring{$[E_7\times SU(2)]^2$}{[E7 x SU(2)]2} theory}
\label{sec:so24so8}

Let us now consider the $SO(24)\times SO(8)$ theory. It arises from the ouroboric variant (a) in Figure \ref{fig:ouroboric-variants} (see the capacitor diagram in Figure \ref{fig:capacitors}a, specifically (a1) in Figure \ref{fig:dressed-capacitors}), which corresponds to the $[E_7\times SU(2)]^2$ heterotic theory in the E-limit, see section \ref{sec:e7su2e7su2}. The configuration has two detached boundaries, each with an O8 and 12 D8s, and two unorientifolded identification points, each with 4 D8s (2 and their 2 orientifold images). 

Let us recall the gauge group structure of the configuration before the E-type limit:
\begin{equation}
 SO(12)\times U(1)\times SU(2)\times SO(12)\times U(1)\times SU(2) \mperiod
 \label{e7su2-Iprime}
 \end{equation}
Using rule \ref{d-rule-so}, in the D-type limit where the ouroboros circle shrinks, the two $SO(12)$ factors combine into an $SO(24)$. Then, using rule \ref{d-rule-u}, the two $SU(2)\times U(1)$ factors combine into an $SO(8)$ (note that in this case, the $U(1)$'s are not unpaired, so the remark in rule \ref{d-rule-so} does not apply). 

Let us now consider the matter content in the configuration. Let us start with  the tachyons. In the $[E_7\times SU(2)]^2$ theory, they transform in the $({\bf 1},{\bf 2};{\bf 1},{\bf 2})$; hence, under the group \eqref{e7su2-Iprime}, they transform as singlets of the $SO(12)$'s and in the $({\bf 2},{\bf 2})$ of the $SU(2)$'s. According to rule \ref{d-rule-matter}, the number of tachyons should double in the D-limit, thus becoming 8. To understand in which representation they transform, note the decomposition
\begin{eqnarray}
SO(8) & \quad \quad \rightarrow & \quad \quad \quad  SU(2)^4 \quad \quad \quad\quad \rightarrow\quad\quad \quad  SU(2)^2\nonumber\\
{\bf 8}_v & \quad \rightarrow & \quad ({\bf 2},{\bf 2};{\bf 1},{\bf 1})+({\bf 1},{\bf 1};{\bf 2},{\bf 2})  \quad \rightarrow \quad ({\bf 2},{\bf 2})+4({\bf 1},{\bf 1})\, ,
\end{eqnarray}
where the second arrow is a projection onto the first two $SU(2)$'s. The states that are present in the theory can therefore enhance, using rule \ref{d-rule-matter}, to tachyons in the $({\bf 1}, {\bf 8}_v)$ of the $SO(24)\times SO(8)$.

Consider now the massless fermions. In the $[E_7\times SU(2)]^2$ theory, they transform in the $({\bf 56},{\bf 2};{\bf 1},{\bf 1}) +({\bf 56},{\bf 1};{\bf 1},{\bf 2})\,+({\bf 1},{\bf 2};{\bf 56},{\bf 1})\,+({\bf 1},{\bf 1};{\bf 56},{\bf 2})$. Under the decomposition $E_7\to SO(12)\times SU(2)$ (not to be confused with the above $SU(2)$ factors), we have that ${\bf 56}\to ({\bf 12},{\bf 2})+({\bf 32}, {\bf 1})$, and therefore one realizes that the latter spinor arises from D0-brane states and should be dropped according to rule \ref{d-rule-d0s}. Hence, in terms of the $SO(12)$ subgroup, we have ${\bf 56}\to 2\cdot{\bf 12}$. According to rule \ref{d-rule-matter}, these $4\times 2\times 24=192$ states should double to $384$. As with the tachyons, it is possible to argue how they will be reorganized in representations of the D-limit gauge groups.
In fact, the total fermion content under \eqref{e7su2-Iprime} is two copies of $({\bf 12},{\bf 2};{\bf 1},{\bf 1})\, +({\bf 12},{\bf 1};{\bf 1},{\bf 2})\,+({\bf 1},{\bf 2};{\bf 12},{\bf 1})\,+({\bf 1},{\bf 1};{\bf 12},{\bf 2})$ and one has the decomposition and projection
\begin{eqnarray}
SO(8) & \quad \quad \rightarrow & \quad \quad \quad  SU(2)^4 \quad \quad \quad\quad \rightarrow\quad \quad\quad  SU(2)^2\nonumber\\
{\bf 8}_s & \quad \rightarrow & \quad ({\bf 2},{\bf 1};{\bf 2},{\bf 1})+({\bf 1},{\bf 2};{\bf 1},{\bf 2})  \quad \rightarrow \quad 2\,[\,({\bf 2},{\bf 1})+({\bf 1},{\bf 2})\,]\mperiod
\end{eqnarray}
The massless fermions can then be enhanced to $({\bf 24},{\bf 8}_s)$ of the $SO(24)\times SO(8)$. This accounts for 192 states, but from rule \ref{d-rule-matter} there should be 192 more.
In fact, the massless sector of the heterotic theory also contains fermions in the $({\bf 24},{\bf 8}_v)$, which must be part of the additional massless states arising in the D-limit, completing the expectation from rule~\ref{d-rule-matter}. It is fortunate that the states filling out the $({\bf 24},{\bf 8}_s)$ are already present in the theory before the D-limit, as they would not have been easy to generate in a D-limit enhancement (since spinor representations typically arise from D0-branes, which are not expected to produce massless states in D-limits; see rule~\ref{d-rule-d0s}).

Finally, the global structure of the gauge group, $(Spin(24)\times Spin(8))/\Z_2$, (see \eqref{global}), arises as follows. The massive D0-brane states and the massless ${\bf 8}_s$ that we just discussed produce massive spinor representations, thus requiring the $SO$ algebras to arise from $Spin$ groups. Then, the $\Z_2$ quotient, which requires that vectors of $SO(24)$ must come correlated with $({\bf 8}_v$ or ${\bf 8}_s$) representations of $SO(8)$, arises from the quantum superposition $\Z_2$ in the $[E_7\times SU(2)]^2$ theory. This requires the  representations of the four factors (${\bf 56}$'s of $E_7$ or ${\bf 2}$'s of $SU(2)$) to come in pairs.

\subsubsection{The \texorpdfstring{$U(16)$}{U(16)} from the \texorpdfstring{$(E_8)_2$}{(E8)2} theory}
\label{sec:u16}

Let us now consider the $SU(16)\times U(1)$ theory. It arises from the ouroboric variant (a) in Figure \ref{fig:ouroboric-variants} (see the capacitor diagram in Figure \ref{fig:capacitors}a, and concretely that in Figure \ref{fig:dressed-capacitors}a2), which leads to the $(E_8)_2$ heterotic in the E-limit; see section~\ref{sec:e82}. The configuration has two detached boundaries, each with an O8 and 14 D8s, and two unorientifolded identification points, each with 2 D8s (one and its orientifold image). 

Recall the gauge group structure in the configuration before the E-type limit. We naively have $[SO(14)\times U(1)]^2$, but the quantum superposition $\Z_2$ projects down to the diagonal combination $SO(14)\times U(1)$. This implies that we may use rule \ref{d-rule-so} with the interpretation that this $U(1)$ is unpaired, thus arguing that the D-limit enhanced version of the parent $[SO(14)\times U(1)]^2$ theory has gauge group $SO(32)$. However, we must still impose the  quantum superposition $\Z_2$. Note that the result of this projection in the D-limit need not be an automorphism of the enhanced gauge group, so we need alternative arguments to find the resulting symmetry. A useful guide is that it should correspond to a $\Z_2$ projected version of $SO(32)$ which describes an enhancement of the diagonal $SO(14)\times U(1)$, or, even more simply, the $SO(16)$ in which it is embedded. A simple choice\footnote{Morally, the original $E_8\times E_8$ symmetry before the quantum superposition $\Z_2$ breaks down to a $SO(16)^2$, which enhances to $SO(32)$ in the D-limit, and is ultimately projected down to $U(16)$ by the quantum superposition $\Z_2$.} is $U(16)$, which is moreover very well motivated by the fact that it would correspond to the only D-type heterotic missing in our previous dualities. 

We will now show that the matter content also fits nicely with the proposal that the enhanced gauge group is $U(16)$.

Consider the light matter spectrum. The $(E_8)_2$ theory has a tachyon singlet, already present in the theory before the E-limit, from the D8s at the unorientifolded identification points. 
From rule \ref{d-rule-matter}, this must double in the D-limit, thus completing the two singlet tachyons of $SU(16)$ in the final heterotic theory.
Regarding the fermions, the two ${\bf 248}$'s of the $(E_8)_2$ theory give rise to two ${\bf 120}$'s of the diagonal $SO(16)$, which, upon receiving extra partners in the D-limit enhancement, double their number due to rule \ref{d-rule-matter} and reproduce the $2 ({\bf 120}+{\bf {\overline{120}}})$ of $SU(16)$. Hence, the D-limit of the $(E_8)_2$ theory corresponds to the $SU(16)\times U(1)$ theory, see Table \ref{table:heterotics}.

Admittedly, the derivation of the D-limit in this case relies on a somewhat heuristic application of our rules. This is arguably due to the subtleties in properly interpreting the quantum superposition $\Z_2$, especially away from the E-limit where it admitted a sharper definition. It is plausible that a more careful understanding of its action would also explain the global structure of the gauge group in this theory, see \eqref{global}.

\subsubsection{The \texorpdfstring{$SO(16)\times SO(16)$}{SO(16) x SO(16)} heterotic}
\label{sec:so16so16-d}

We have obtained all the D-type 10d non-supersymmetric heterotic theories from D-limits of ouroboros configurations underlying the E-type heterotics. Yet, there is another E-type heterotic theory: the $SO(16)\times SO(16)$ string. It makes sense to ask the question of what the D-limit of the ouroboros configuration underlying the $SO(16)\times SO(16)$ theory is. We now argue that the answer is that one recovers the $SO(16)\times SO(16)$ theory itself. This fits well with the idea that this theory plays a central role, and that it can be regarded as belonging to both E- and D-type heterotics.

Consider the ouroboric variant (c) in Figure \ref{fig:ouroboric-variants} (see the capacitor diagram in Figures \ref{fig:capacitors}c, \ref{fig:dressed-capacitors}c), which leads to the $SO(16)\times SO(16)$ heterotic theory in the E-limit, c.f.~section~\ref{sec:so16so16-e}. The configuration has two glued boundaries, each with an O8 and 16 D8s.

In the D-limit, the gauge group is not enhanced because rule \ref{d-rule-so} does not apply to two glued boundaries. Therefore, the gauge symmetry is still $SO(16)\times SO(16)$. Regarding the matter content, the initial theory does not contain a tachyon, so neither does the final theory after the D-limit. This agrees with rule \ref{d-rule-matter}, as $2\times 0=0$. As for the fermions, the states in the $({\bf 16},{\bf 16})$ arise from the sector between the two boundaries and, according to rule \ref{d-rule-matter}, they survive as states in the $({\bf 16},{\bf 16})$ after the D-limit. On the other hand, the states in the ${\bf 128}$ of the $SO(16)$'s arise from D0-branes, so they are naively not present in the limit. Therefore, we obtain 256 fermions from the $({\bf 16},{\bf 16})$. Using rule \ref{d-rule-matter}, there should be 256 more states in the D-limit, which indicates that the theory we are recovering in the D-limit is an $SO(16)\times SO(16)$ heterotic. In fact, a possibility is that in this special case, the generically massive D0-brane states become massless in the D-limit as well and restore the ${\bf 128}$ of the $SO(16)$'s in the final theory, thus generating the missing 256 states and reproducing the spectrum in Table \ref{table:heterotics}. The fact that our rules fail to account for these spinor states in the D-limit is not a drawback, but it rather highlights the central role of the $SO(16)\times SO(16)$ heterotic theory and hints at remarkable self-duality properties in the non-supersymmetric duality web.

\medskip

This concludes our discussion about how considering the D-limits of ouroboric variant configurations (morally, shrinking their circle to zero length) reproduces the different D-type heterotics using a simple set of rules. In a precise sense, we have explained the quantum geometry and the physics underlying the choice of Wilson lines required to carry out the proposal in \cite{Baykara:2026vdc} to relate E- and D-type heterotics, allowing us to provide a complete characterization of the different heterotic theories, their light spectra, and the global structure of their gauge groups. We expect that these relations will yield a rich web of dualities involving the heterotic theories in the future.

As in the case of the E-limit, the rules that we have used in the D-limits are heavily inspired by the standard ones in geometric contexts. It would be interesting to gain some insight into why they work so well in the non-geometric (and non-supersymmetric) setups.

\begin{figure}[htb]
\begin{center}
\includegraphics[scale=.35]{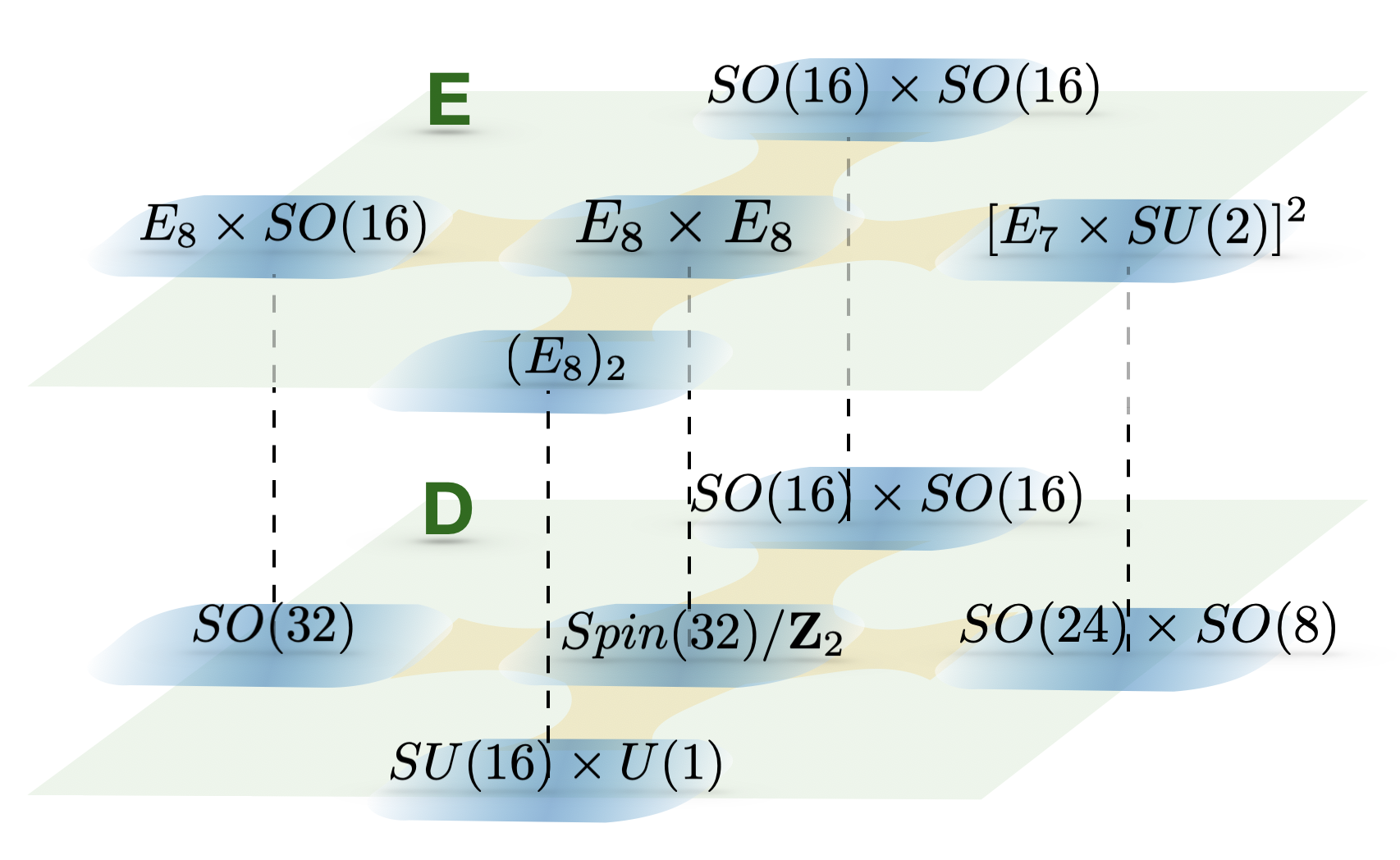}
\caption{\small The different E- and D-type 10d supersymmetric and non-supersymmetric heterotic theories, with their relations via going up and down the E- and D-limits.}
\label{fig:relations}
\end{center}
\end{figure}

Finally, we would like to mention that the relations between the different 10d supersymmetric and non-supersymmetric heterotics of E- and D-type, and the central role of the $SO(16)\times SO(16)$ theory, were analyzed in \cite{BoyleSmith:2023xkd} using the modern understanding of fermionization in 2d. The picture that we have developed leads to a series of connections that bear a striking resemblance, albeit from a quantum-geometric viewpoint. We illustrate these relations in Figure \ref{fig:relations}, which may be regarded as a 3d version of Figure 3 in \cite{BoyleSmith:2023xkd} upon folding it, using the $SO(16)\times SO(16)$ theory as the pivotal hinge.

\section{Heterotic junctions and bouquets}
\label{sec:junctions}

In this section, we employ the quantum-geometric picture that we have developed in the previous sections for the construction of the recently introduced junctions of 10d heterotic string theories \cite{Altavista:2026edv} (see also \cite{Tachikawa:2026jaj}, and see \cite{Hellerman:2010dv} for an early reference). These realize multi-pronged cobordism relations among 10d  theories, in the spirit of the string cobordism conjecture \cite{McNamara:2019rup}, and generalize the concept of end-of-the-world boundaries (see~\cite{Horava:1995qa,Horava:1996ma,Polchinski:1995df}\footnote{See \cite{Buratti:2021yia,Buratti:2021fiv,Angius:2022aeq,Blumenhagen:2022mqw,Angius:2022mgh,Angius:2023xtu,Blumenhagen:2023abk,Angius:2023uqk,Angius:2024zjv,Delgado:2023uqk, Antonelli:2019nar} for other dynamical realizations of cobordism ETW boundaries in diverse contexts.}) or interface domain walls between two theories (see \cite{Heckman:2025wqd,Anastasi:2026cus,Torres:2026vxx} for recent studies of the IIA/IIB interface). Therefore, they provide an interesting tool to explore general cobordism relations between different 10d theories.

Junctions in \cite{Altavista:2026edv,Tachikawa:2026jaj} were explicitly built using a worldsheet perspective, analogous to the realization of string cobordisms by moving up and down the RG flow in the spirit of \cite{Gaiotto:2019asa}. Our approach is complementary, as it is based on the quantum-geometric realization underlying the relevant theories (and is hence reminiscent of \cite{Hellerman:2010dv}).

Although any junction is topologically allowed by a sufficiently general cobordism (because all theories are in the trivial class, according to the cobordism conjecture),  a particular class of junctions of chiral 10d theories, dubbed bouquets, was highlighted in \cite{Altavista:2026edv}, in which chiral fields of each theory, rather than being simultaneously gapped upon reaching the junction\footnote{These symmetric mass generation mechanisms require strong coupling dynamics and are unknown in 10d theories; see \cite{Angius:2024pqk} for mechanisms in 6d and 4d chiral gravitational theories, and, e.g., \cite{Razamat:2020kyf,Tong:2021phe,Wang:2022ucy} for 2d and 4d QFT examples.}, flow across it into another branch (in a bouquet of chiral flow-ers). We will use our techniques to recover known bouquets \cite{Altavista:2026edv,Tachikawa:2026jaj} and to propose new ones. Ultimately, our findings are exploratory and further evidence is needed for the new bouquets that we propose.

\subsection{Junctions of heterotic geometries and trefoil diagrams}
\label{sec:junctions-ide}

The general strategy to construct junctions using the geometric realization of heterotic theories is fairly simple. We glue different ouroboros and/or geometric type I' configurations and take the E- or D-limits to obtain 10d theories. Note that in our discussion we simply consider the properties of junctions at the topological level, without trying to find a dynamical realization satisfying the equations of motion in spacetime.

\subsubsection{The trivalent bouquet of the two 10d supersymmetric heterotics}
\label{sec:bouquet-guay}

\begin{figure}[htb]
\begin{center}
\includegraphics[scale=.3]{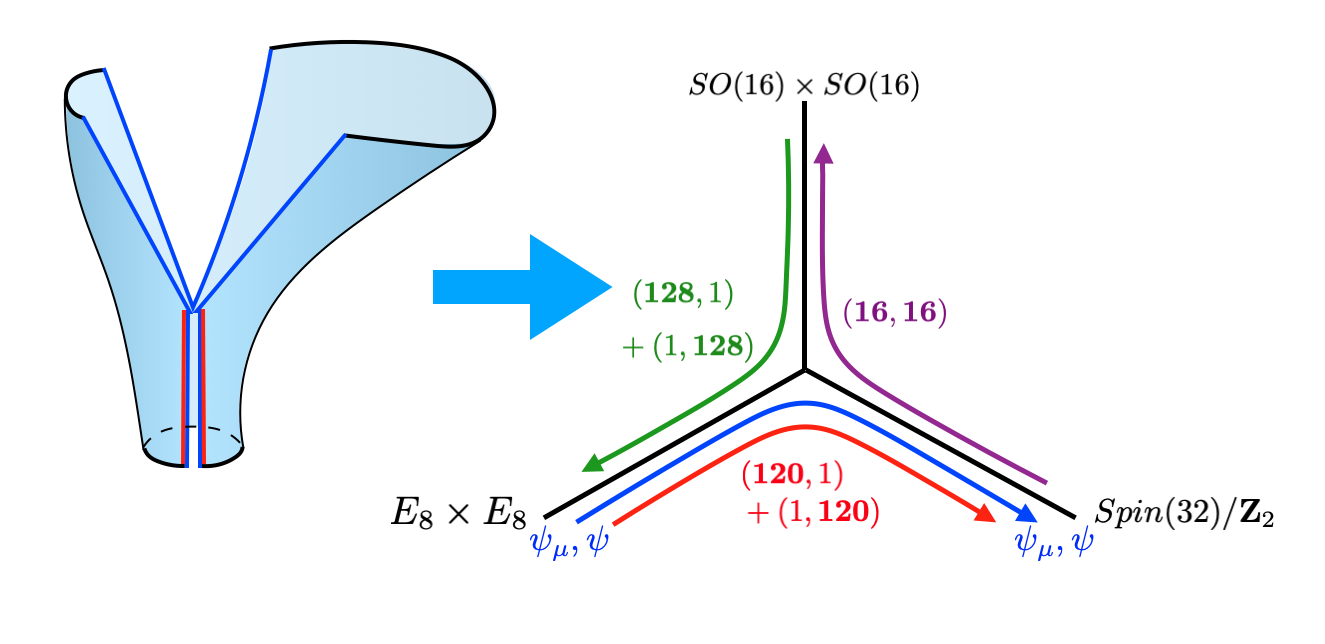}
\caption{\small a) Geometric realization of the bouquet of the $E_8\times E_8$, the $Spin(32)/\Z_2$ and the $SO(16)\times SO(16)$ heterotics. b) The chiral flow across the bouquet, as explained in \cite{Altavista:2026edv,Tachikawa:2026jaj}.}
\label{fig:junction-guay}
\end{center}
\end{figure}

For example, let us consider the bouquet of the $E_8\times E_8$, $Spin(32)/\Z_2$, and $SO(16)\times SO(16)$ theories proposed in \cite{Altavista:2026edv} (based on relations among these theories pointed out in \cite{Basile:2023knk,Tachikawa:2024ucm}) and built in \cite{Tachikawa:2026jaj}. The 10d supersymmetric $E_8\times E_8$ and $Spin(32)/\Z_2$ branches can be built by considering geometric type I' configurations in the E- and D-limits, respectively, while the $SO(16)^2$ theory can be realized by a quantum ouroboros of type (c) in Figure \ref{fig:ouroboric-variants}. A depiction of a junction describing the gluing of these geometries is shown in Figure \ref{fig:junction-guay}a. 

Note that, since the realization of the $SO(16)^2$ theory is in terms of a quantum geometry, the nature of the junction is quantum-geometric as well. In particular, the different boundaries of the branches all coincide at a point in the configuration. This is important to give rise to a realization of the chiral flow of the bouquet, shown in Figure \ref{fig:junction-guay}b, as follows. The fermions in the $({\bf 16},{\bf 16})$ of the $SO(16)\times SO(16)$ theory, arising from open strings stretching between the two glued boundaries of the ouroboros, turn into the gaugini in the $({\bf 16},{\bf 16})$ of $SO(16)\times SO(16)\subset SO(32)$ in the $Spin(32)/\Z_2$ theory. The latter states arise from open strings stretching across the geometric type I' interval of the $Spin(32)/\Z_2$ theory, which shrinks to zero size in its D-limit. The fermions in the $({\bf 128},{\bf 1})+({\bf 1},\bf 128)$ of the $SO(16)^2$ theory, arising from D0s at the two boundaries of the ouroboros, turn into the gaugini in the $({\bf 128},{\bf 1})+({\bf 1},\bf 128)$ of the $SO(16)^2\subset E_8\times E_8$. The latter states arise from D0s at the boundaries of the type I' interval in the $E_8\times E_8$ branch, which become massless in the corresponding E-limit. Finally, the gaugini in the $({\bf 120},{\bf 1})+({\bf 1},{\bf 120})$ of $SO(16)\times SO(16) \subset SO(32)$ flow to similar gaugini of $SO(16)\times SO(16) \subset E_8\times E_8$.

We stress again that the joining of boundaries at the junction should be regarded as a quantum superposition of possible classical gluings. For instance, the flow of the $({\bf 16},{\bf 16})$ from the $SO(16)\times SO(16)$ to the $Spin(32)/\Z_2$ branch requires that the boundaries of the two theories be continuously joined; at the same time, the flow of the $({\bf 128},{\bf 1})+({\bf 1},\bf 128)$ from the $SO(16)\times SO(16)$ to the $E_8\times E_8$ branch requires that the boundaries of the two theories be continuously joined; finally, the flow of the $({\bf 120},{\bf 1})+({\bf 1},{\bf 120})$ from the $Spin(32)/\Z_2$ to the $E_8\times E_8$ branch requires that the boundaries of these two theories be continuously joined. All possibilities must coexist in superposition for a consistent chirality flow. In practice, this is analogous to fixing a classical joining of boundaries and allowing some degrees of freedom (e.g., D0-branes) to jump between disconnected boundaries in that classical gluing. This will be an important ingredient in the construction of junctions of non-supersymmetric heterotics, to which we now turn.

\subsubsection{Junctions of non-supersymmetric heterotics and trefoil diagrams}

\begin{figure}[htb]
\begin{center}
\includegraphics[scale=.35]{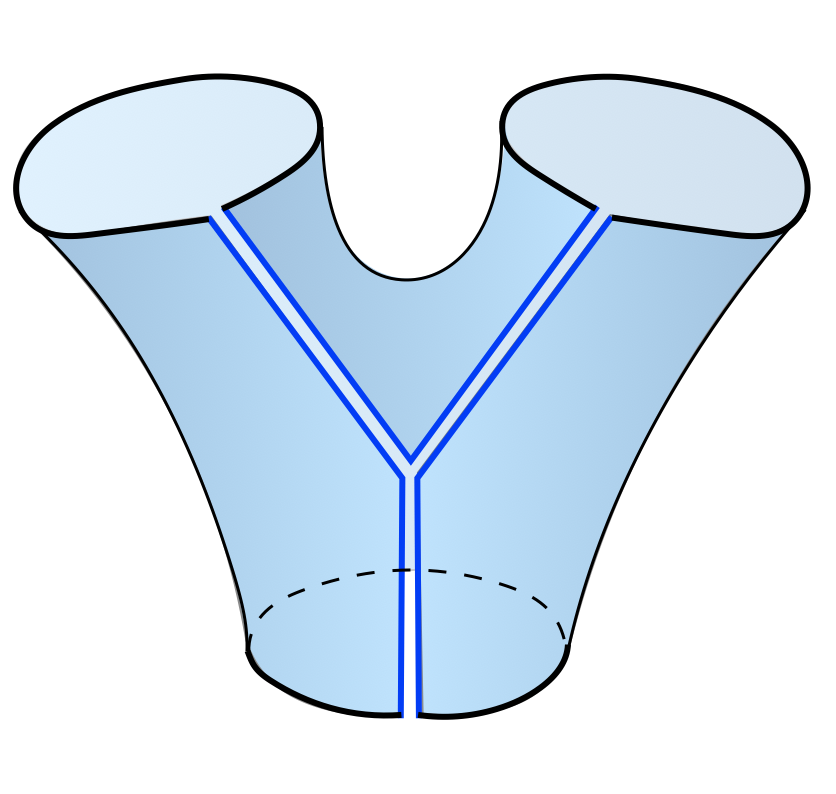}
\caption{\small Geometric realization of a generic 3-branch junction of non-supersymmetric heterotic theories.}
\label{fig:pants}
\end{center}
\end{figure}

One can similarly construct junctions of non-supersymmetric heterotic theories by considering pair of pants diagrams interpolating between the corresponding ouroboros configurations (and subsequently taking the E- and D-limits, or combinations thereof), see Figure \ref{fig:pants}. We will focus on 3-branch junctions only; the generalization of the key ideas to junctions with more branches is straightforward.

\begin{figure}[htb]
\begin{center}
\includegraphics[scale=.32]{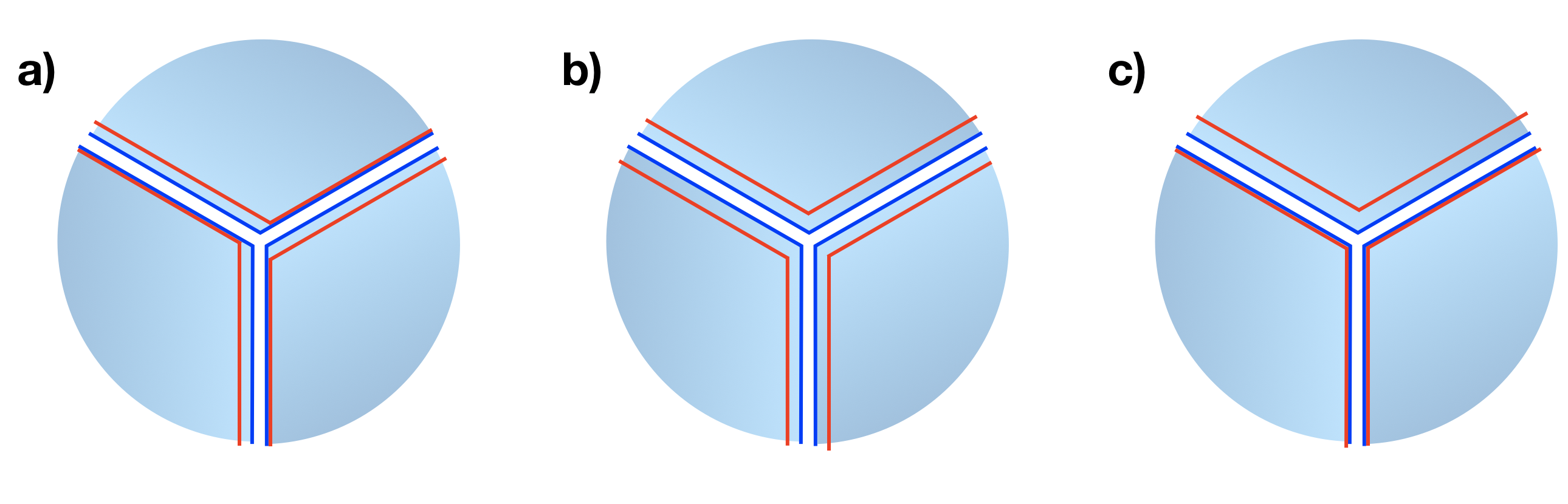}
\caption{\small  Examples of trefoil diagrams describing 3-branch junctions of non-supersymmetric heterotic theories.}
\label{fig:trefoils}
\end{center}
\end{figure}

Actually, Figure \ref{fig:pants} corresponds to the depiction of the quantum geometry using the unresolved picture of the heterotics \cite{Baykara:2026vdc}. Therefore, we need to promote this picture by dressing the gap region in each asymptotic branch with the ingredients of the corresponding capacitor diagrams from Figure \ref{fig:dressed-capacitors} (the identification points and the D8 locations) to turn it into an ouroboric variant. Clearly, given the locality of the configuration around the capacitor diagram, the topology of the junction is encoded in the recombination of ingredients in the region where the gaps join. By zooming into that region, we can encode the information of the junction in simple diagrams, which we refer to as trefoil diagrams; see some examples in Figure \ref{fig:trefoils}.

A general remark is that, in analogy with the bouquet of section \ref{sec:bouquet-guay}, the classical joining of boundaries should be promoted to a quantum superposition of joining patterns. Alternatively, we may keep the depicted classical joining pattern but allow degrees of freedom on the boundaries to `jump' across the gap.

\subsection{Examples of junctions of E-type heterotics}
\label{sec:junctions-e-examples}

A general question is whether any given trefoil diagram describes a physical bouquet, in which the chiral flow is determined by the joining of boundaries and identification lines. In this section, we provide several explicit examples where this is the case, but also others in which a given trefoil diagram does not describe a consistent bouquet. 

Consider the trefoil diagram in Figure \ref{fig:trefoils}a, in which all three branches are in the E-limit. This describes a bouquet of three $E_8\times SO(16)$ theories, which was, in fact, constructed in \cite{Altavista:2026edv} using a different technique. We now show how to easily recover the corresponding chiral flow. The chiral matter in each of the $E_8\times SO(16)$ theories is given by the chiral fermions in the spinor representation of $SO(16)$. In our ouroboros realization, these arise from D0-brane states at the glued boundary (i.e., the blue line coinciding with the red line of identification points). Taking into account that D0s are allowed to jump across the gap due to the quantum superposition of joining patterns mentioned above, it follows that each spinor of $SO(16)$ can turn into a spinor of the $SO(16)$ on any of the two other branches, as explained in \cite{Altavista:2026edv}.

Let us consider another example. Take the trefoil diagram in Figure \ref{fig:trefoils}b, which corresponds to a junction of three theories with two detached boundaries. By placing 12 D8s on all the boundaries and taking the E-limit in all three branches, we get a junction of three $[E_7\times SU(2)]^2$ theories. The chiral matter in each of these theories is the set of fermions in the $({\bf 56},{\bf 2};{\bf 1},{\bf 1})_++ ({\bf 56},{\bf 1};{\bf 1},{\bf 2})_-+({\bf 1},{\bf 2};{\bf 56},{\bf 1})_- +({\bf 1},{\bf 1};{\bf 56},{\bf 2})_+$. Let us construct a consistent chiral flow for this theory inspired by the trefoil diagram in Figure \ref{fig:trefoils}b. Consider the $({\bf 56},{\bf 2};{\bf 1},{\bf 1})_+$ in the bottom branch in the diagram, which arises from the two gauge factors on the left side of the corresponding gap. These propagate continuously along the lower-left wedge of the trefoil to the upper-left branch, hence turning into a $({\bf 56},{\bf 2};{\bf 1},{\bf 1})_+$ of that $[E_7\times SU(2)]^2$. A similar pattern applies to the $({\bf 1},{\bf 1};{\bf 56},{\bf 2})_+$, which propagates smoothly to the upper-right branch. On the other hand, the $({\bf 56},{\bf 1};{\bf 1},{\bf 2})_-$ of the bottom branch is charged under gauge factors on different sides of the gap. Upon reaching the junction, the two sides separate, with the ${\bf 56}$ moving along the lower-left wedge into the upper-left branch and the ${\bf 2}$ moving along the lower-right wedge into the upper-right branch. One possibility, if there is a quantum superposition of joining patterns, is that the ${\bf 56}$ may be allowed to jump to another boundary, e.g., the upper boundary of the upper-right branch, so that the state turns into a $({\bf 56},{\bf 1};{\bf 1},{\bf 2})_-$ of that branch. Alternatively, keeping the classical joining pattern in the picture, the initial $({\bf 56},{\bf 1};{\bf 1},{\bf 2})_-$ of the bottom branch may reach the junction, emit a tachyon in the $({\bf 1},{\bf 2};{\bf 1},{\bf 2})$ of the upper-right branch, and turn into a  $({\bf 56},{\bf 1};{\bf 1},{\bf 2})_-$ of the upper-left branch. 

It is clear that one can obtain a consistent chiral flow for all chiral fermions in the model, suggesting that the trefoil diagram in Figure \ref{fig:trefoils}b describes a consistent bouquet of three $[E_7\times SU(2)]^2$ theories. It would be interesting to construct this configuration using worldsheet techniques.

Finally, let us show that there exist trefoil diagrams which however cannot describe a consistent chiral flow for the junction.  Consider the trefoil in Figure \ref{fig:trefoils}c, with the three branches in the E-limit. This would naively describe a junction of one $SO(16)\times SO(16)$ and two $E_8\times SO(16)$ theories. However, it is easy to see that it is not possible to construct a consistent chiral flow among these theories. For instance, the fermions in the $({\bf 16},{\bf 16})_-$ of $SO(16)\times SO(16)$ have no appropriate transformation channel upon reaching the junction, because the only fermions in the two other theories are in spinor representations of their corresponding $SO(16)$'s.
This is an illustrative example that generic trefoil diagrams need not produce consistent physical bouquets (i.e., configurations with consistent chiral flows), although they may still define consistent junctions if they provide a mechanism to gap the chiral fermions at the junction. Understanding the precise constraints (and their quantum-geometric origin) that trefoils must satisfy to produce consistent bouquets, or to lead to consistent junctions via gapping of fermions, is an interesting question that we leave for future work.

\subsection{Examples of junctions of D-type heterotics}
\label{sec:junctions-d-examples}

Let us now consider some examples of seemingly consistent bouquets that can be obtained by trefoil diagrams with theories in the D-limit. 

Consider the trefoil diagram in Figure \ref{fig:trefoils}a, in which all three branches are in the D-limit. Using our results from section \ref{sec:d-limit}, this describes a junction between three $SO(32)$ theories (which was constructed in \cite{Altavista:2026edv}). Since these theories are non-chiral (actually, purely bosonic at the massless level), there is no non-trivial chiral flow.

Consider now the trefoil diagram in Figure \ref{fig:trefoils}b, in which all boundaries have 12 D8s, and take all three branches in the D-limit. The configuration becomes a junction of three $SO(24)\times SO(8)$ theories. In this case, a non-trivial chiral flow can be obtained by adapting the chiral flow of the same diagram in the E-limit. The result is that the fermion in the $({\bf 24},{\bf 8}_s)$ in one branch can turn into the analogous field in another branch by the emission of a tachyon in the $({\bf 1},{\bf 8}_v)$ in the third branch. On the other hand, the fermion in the $({\bf 24},{\bf 8}_v)$ of a branch should be able to propagate into the analogous field in any other branch, although this is harder to assess as these fields are not present in the E-limit; hence, we do not have their detailed microscopic description. Overall, this seems to produce a consistent bouquet, and it would be interesting to confirm this using independent techniques. 

Finally, let us consider the trefoil in Figure \ref{fig:trefoils}c, with all three branches in the D-limit. The configuration would describe a junction of one $SO(16)\times SO(16)$ and two $SO(32)$ theories. Since the latter are non-chiral (in fact, purely bosonic at the massless level), the configuration cannot admit a consistent chiral flow for the chiral matter in the $SO(16)\times SO(16)$ branch. This provides an even simpler explanation of why the E-limit of this trefoil failed to provide a consistent bouquet of one $SO(16)\times SO(16)$ and two $E_8\times SO(16)$ theories. In any case, this proves once again that a generic trefoil diagram does not imply the existence of a consistent bouquet. We leave the exploration of trefoil geometries that do not produce bouquets and their possible interpretation in terms of junctions with symmetric mass generation mechanisms for future research.

\section{Conclusions}
\label{sec:conclusions}

We have found that M-theory on the $\Z_2$ quotient of $\8$ that exchanges the two $\S^1$'s generates the 10d non-supersymmetric heterotic string theories, as envisioned in~\cite{Baykara:2026vdc}. More importantly, our analysis, based on gauge enhancements in type I' string theory, sheds light on how to work with degrees of freedom living at the boundaries in these non-geometric compactifications. Our microscopic understanding relies on a set of rules that mimic those of standard D-branes and O-planes in string theory, with some crucial differences that should be the hallmark of the $\8$ setup. Some of these rules are ad hoc prescriptions that will probably admit more refined formulations---rule~\ref{d-rule-matter} in the D-limit case is perhaps the most transparent example of this---but it is encouraging that, in this strikingly simple framework, non-supersymmetric string theories seem to arise from M-theory.

Several open questions remain. The most pressing one is perhaps how to include the massive string states in the formalism. This question is relevant not only in the heterotic case but also in the type 0 cases of~\cite{Baykara:2026gem,Altavista:2026evd}. If non-supersymmetric string theories actually stem from the moduli space of M-theory on these types of non-geometric setups, there must be checks that one can perform to test the proposal against the ultimate consistency of string theory. We found a first hint of this in section~\ref{sec:d-heterotics}, with the appearance of non-supersymmetric D0-brane states in D-limits.

Another mystery that remains to be unraveled is the bound on the number of D8-branes at unorientifolded identification points of section~\ref{sec:e-microscopic}. As we mentioned there, this number must be at most $2+2$ orientifold images to yield proper 10d heterotic theories. This is reminiscent of the bound on D-brane charges in WZW models and can be formulated by postulating a version of an exclusion principle at the unorientifolded identification points. However, there is no clear reason for this to be the case; more work is required to gain further insight into the issue.
In practice, this is equivalent to asking why the E-limits and the D-limits are the only ways to get 10d heterotic theories out of ouroboros. In fact, other possibilities may be realized as lower-dimensional heterotic strings at enhancement points (see~\cite{Fraiman:2023cpa}), and the missing ingredient would be an obstruction, in the $\8$ setup, for ouroboric variants to uplift to 10d.

On top of this, it would be interesting to understand the enhancements---in particular, the D-limits that are less transparent in the type I' ouroboros picture---from the perspective of non-supersymmetric heterotic strings on $\S^1$, using the tools of~\cite{Fraiman:2023cpa}. This could provide new hints about the dynamics of the ouroboros configurations and, at the same time, uncover strongly coupled phenomena associated with maximal enhancements, for example, along the lines of~\cite{Morrison:1996xf}.

A point that remains only partially understood is the structure of the junctions in the ouroboros picture. In section~\ref{sec:junctions}, we have found that the ouroboros setup provides an elegant pictorial way of generating junctions. However, some of the junctions that seem available are not consistent and cannot accommodate the chiral flow (unless other mechanisms for gapping fermions are available).
In fact, what is probably lacking is a rationale to distinguish junctions that can have an ouroboric origin from those that cannot. 
Moreover, within the compactification of M-theory on $\8$, a natural question is whether one can find junctions with open strings, using the type 0 embeddings of~\cite{Baykara:2026gem,Altavista:2026evd}. This is an interesting question that we leave for future work.

Despite being in an early stage, the picture that emerges from M-theory on quantum geometries from $\8$ is that non-supersymmetric strings can be associated with points in this non-geometric moduli space.
However, this is only part of the issue with non-supersymmetric string theories. Understanding how tachyons and tadpole instabilities from string theory manifest themselves in the M-theory framework is an equally important topic, which would clarify how the dynamics of non-supersymmetric gravity is encoded in these non-geometric setups within M-theory.

\section*{Acknowledgements}

We are pleased to thank Edoardo Anastasi, Roberta Angius, and Bernardo Fraiman for useful conversations.
This work is supported by the grants CEX2020-001007-S, PID2021-123017NB-I00, PID2024-156043NB-I00, funded by MCIN\slash AEI\slash10.13039\slash501100011033, and ERDF A way of making Europe. The work of C.A. is supported by the fellowship LCF/BQ/DFI25/13000111 from ``La Caixa'' Foundation (ID 100010434). S.R. is supported by the ERC Starting Grant QGuide101042568 - StG 2021. C.W. is supported by program PIPF-2024/TEC-34293 from Comunidad de Madrid.

\appendix

\section{D8/O8 systems and exceptional symmetry enhancement}
\label{app:enhancement}

In this appendix, we briefly review the non-perturbative gauge symmetry enhancements arising in type I' configurations of D8s and O8s \cite{Polchinski:1995df}, uncovered in \cite{Seiberg:1996bd} from the study of SCFT fixed points of 5d gauge theories realized on the worldvolume of D4-brane probes of the type I' configuration (see \cite{Bergman:1997py,Bachas:1997kn} for a direct discussion using D0-brane dynamics in the O8/D8 system). The enhancement of gauge symmetries on the D8s to exceptional groups is reflected in the corresponding enhancement of global symmetry of the 5d theory as the strongly coupled SCFT is reached. The physical interpretation of the exceptional symmetry arises from the M-theory lift of the O8/D8 system to a Ho\v{r}ava--Witten wall, with the $E_8$ gauge symmetry possibly partially broken by Wilson lines.

We consider type I' string theory on an interval $\S^1/\Z_2$, bounded by two O8-planes. The background includes $16$ D8s distributed along the interval. We consider a  D4-brane probe extended along the non-compact directions and located near the O8-plane, whose worldvolume theory realizes a 5d $\mathcal{N}=1$ $SU(2)$ supersymmetric theory, whose Coulomb branch is parametrized by the D4-brane position $\phi$ along the interval.

When $N_f$ D8s (together with their $N_f$ orientifold images) coincide with an O8-plane, the low-energy theory on the probe is an $SU(2)$ gauge theory with $N_f$ hypermultiplets. As explained in \cite{Seiberg:1996bd}, for $N_f \leq 7$, the strong coupling limit $g \to \infty$ leads to an interacting fixed point with enhanced global symmetry $E_{N_f+1}$.

The key dynamical ingredient is that the effective gauge coupling depends on the position due to the backreaction of the D8s \cite{Polchinski:1995df}:
\begin{equation}
\frac{1}{g_{\text{eff}}^2(\phi)} = \frac{1}{g_0^2} + 2(8 - N_f)\,\phi \mcomma 
\end{equation}
and can diverge at finite $\phi$, signaling a strongly coupled fixed point. The divergence of $g_{\text{eff}}$ corresponds to a divergence of the type I' dilaton near the O8-plane. Physically, this indicates that the naive perturbative description breaks down and new light degrees of freedom appear. In this regime, the system is better described by M-theory on an interval (the Ho\v{r}ava--Witten picture), where the O8-plane corresponds to an $E_8$ boundary, and the D8 positions correspond to Wilson lines breaking $E_8$ to subgroups.

Thus, tuning the coupling to diverge at the O8-plane corresponds to restoring an enhanced $E_n$ symmetry, where $n = N_f + 1$. From the type I' perspective, the additional gauge bosons in the spinor representation of $SO(2N_f)$ arise from states of D0-branes, stuck at the O8-plane, which carry fermion zero modes charged under the D8s in the system.

Let us mention two explicit realizations of this phenomenon, which serve as inspirations in the main text. We focus on the local region near an O8, since the physics of the enhancement is local. The first system corresponds to a configuration with 12 D8s (i.e., $N_f=6$) on top of the O8, with 2 additional D8s (one and its orientifold image) located away from the O8. At weak coupling, the gauge symmetry is $SO(12) \times U(1)$. As the separated D8s are brought toward the O8 and the coupling is tuned so that the dilaton diverges at that location, additional states become massless, leading to the enhancement
\begin{equation}
SO(12)\times U(1) \quad \longrightarrow \quad E_7\mperiod
\end{equation}
From the 5d field theory perspective, this corresponds to the $N_f=6$ theory flowing at strong coupling to the $E_7$ fixed point. 

The second example corresponds to a configuration with 14 D8s on top of the O8, with 2 additional D8s (one and its orientifold image) located away from the O8. At weak coupling, the gauge symmetry is $SO(14) \times U(1)$. As the separated D8s are brought toward the O8 and the coupling is tuned so that the dilaton diverges at that location, additional states become massless, leading to the enhancement
\begin{equation}
SO(14)\times U(1) \quad \longrightarrow \quad E_8 \mperiod
\end{equation}
From the 5d field theory perspective, this corresponds to the $N_f=7$ theory flowing at strong coupling to the $E_8$ fixed point.

As already mentioned, these phenomena are naturally understood in the M-theory framework, where exceptional symmetries arise from the structure of the $E_8$ gauge theories on the Ho\v{r}ava--Witten boundaries.

\bibliographystyle{utphys}
\bibliography{mybib}

\end{document}